\documentstyle[aps,eqsecnum,preprint,floats,epsf,psfig]{revtex}
\begin{document}
\def\be{\begin{eqnarray}}
\def\en{\end{eqnarray}}
\def\up{\uparrow}
\def\dw{\downarrow}
\def\non{\nonumber}
\def\la{\langle}
\def\ra{\rangle}
\def\nc{N_c^{\rm eff}}
\def\vp{\varepsilon}
\def\vma{{_{V-A}}}
\def\vpa{{_{V+A}}}
\def\m{\hat{m}}
\def\fp{{f_{\eta'}^{(\bar cc)}}}
\def\half{{{1\over 2}}}
\def\pr{{\sl Phys. Rev.}~}
\def\prl{{\sl Phys. Rev. Lett.}~}
\def\pl{{\sl Phys. Lett.}~}
\def\np{{\sl Nucl. Phys.}~}
\def\zp{{\sl Z. Phys.}~}
\def\lsim{ {\ \lower-1.2pt\vbox{\hbox{\rlap{$<$}\lower5pt\vbox{\hbox{$\sim$}
}}}\ } }
\def\gsim{ {\ \lower-1.2pt\vbox{\hbox{\rlap{$>$}\lower5pt\vbox{\hbox{$\sim$}
}}}\ } }

\font\el=cmbx10 scaled \magstep2
{\obeylines
\hfill IP-ASTP-01-98
\hfill March, 1998}

\vskip 1.5 cm

\centerline{\large\bf Nonfactorizable Effects in Spectator and Penguin}
\centerline{\large\bf Amplitudes of Hadronic Charmless $B$ Decays}
\medskip
\bigskip
\medskip
\centerline{\bf Hai-Yang Cheng}
\medskip
\centerline{ Institute of Physics, Academia Sinica}
\centerline{Taipei, Taiwan 115, Republic of China}
\medskip
\centerline{\bf B. Tseng}
\medskip
\centerline{Department of Physics, National Cheng-Kung University}
\centerline{Tainan, Taiwan 700, Republic of China}
\bigskip
\bigskip
\bigskip
\centerline{\bf Abstract}
\bigskip
{\small 
  Nonfactorizable effects in hadronic charmless $B\to PP,~VP$ decays can be 
parametrized in terms of the effective number of colors $\nc$ in the
effective parameters $a_i^{\rm eff}$ that are linear combinations of 
Wilson coefficients.
It is shown that $\nc(V+A)$ in the penguin amplitudes induced
by the $(V-A)(V+A)$ four-quark operators is different from $\nc(V-A)$ in
the decay amplitudes arising from the $(V-A)(V-A)$ operators. Central
values of the branching ratios for $B^\pm\to\omega \pi^\pm$ and $B\to\pi\pi$ 
decays favor $\nc(V-A)\approx 2$, in accordance with the nonfactorizable
effect observed in $B\to D^{(*)}\pi(\rho)$. 
Measurements of the interference effects in $B^-\to\pi^-(\rho^-)\pi^0
(\rho^0)$ decays will provide a more decisive test on the parameter $\nc(V-A)$.
However, $\nc(V+A)\sim 2$ is ruled out by $B^\pm\to\phi K^\pm$. We find that
the current bound on $B^\pm\to\phi K^\pm$ implies
$\nc(V+A)\gsim 4.3$, which is subject to the corrections from
$W$-annihilation and space-like penguin effects. With $\nc(V-A)\approx 2$ 
we show that the branching ratio of $B\to\eta' K$ is enhanced considerably at
small values of $1/\nc(V+A)$ so that it is compatible with experiment.
In particular, the measurement of $B^0\to\eta' K^0$ is now well 
explained without resorting to any new mechanism or new physics beyond
the Standard Model. It is crucial to measure the charged and neutral 
decay modes of $B\to\phi K$ 
and $B\to\phi K^*$ to test the generalized factorization hypothesis. Finally,
we point out that it is difficult to understand the observed large branching
ratio of $B^\pm\to\omega K^\pm$ within the present framework. Inelastic
final-state interactions may alleviate the difficulty with this decay mode.

}

\pagebreak
 
\section{Introduction}
To describe the hadronic weak decays of mesons, the mesonic
matrix elments are customarily evaluated under the factorization hypothesis
so that they are factorized into the product of two matrix elements of
single currents, governed by decay constants and form factors. In the
naive factorization approach, the relevant Wilson coefficient functions
for color-allowed external $W$-emission (or so-called ``class-I") and 
color-suppressed (class-II) internal $W$-emission
amplitudes are given by $a_1=c_1+c_2/N_c$, $a_2=c_2+c_1/N_c$, respectively, 
with $N_c$ the number of colors. In spite of its tremendous simplicity, naive
factorization encounters two major difficulties. First, it never works for
the decay rate of class-II decay modes, though it usually operates for
class-I transition. For example, the predicted decay rate of the 
color-suppressed decay $D^0\to\overline{
K}^0\pi^0$ in the naive approach is too small when compared with experiment 
(for a review, see \cite{Cheng89}).
Second, the hadronic matrix element under factorization is
renormalization scale $\mu$ independent as the vector or axial-vector
current is partially conserved. Consequently, the amplitude $c_i(\mu)
\la O\ra_{\rm fact}$ is not truly physical as the scale dependence of
Wilson coefficients does not get compensation from the matrix elements.
The first difficulty indicates that it is inevitable and 
mandatory to take into account nonfactorizable contributions, especially 
for class-II decays, to render the color suppression of internal $W$
emission ineffective. The second difficulty also should
not occur since the matrix elements of four-quark 
operators ought to be evaluated in the same renormalization scheme as that for 
Wilson coefficients and renormalized at the same scale $\mu$. 

   Because there is only one single form factor (or Lorentz scalar) 
involved in the
class-I or class II decay amplitude of $B\,(D)\to PP,~PV$ decays ($P$: 
pseudoscalar meson, $V$:
vector meson), the effects of nonfactorization can be lumped into the
effective parameters $a_1$ and $a_2$ \cite{Cheng94}:
\footnote{As pointed out in \cite{Kamal94}, the general amplitude of 
$B(D)\to VV$ decay consists of three independent Lorentz scalars, 
corresponding to $S$-, $P$- and $D$-wave amplitudes. Consequently, it
is in general not possible to define an effective $a_1$ or $a_2$ unless
nonfactorizable terms contribute in equal weight to all partial wave 
amplitudes.}
\be
a_1^{\rm eff}=c_1+c_2\left({1\over N_c}+\chi_1\right),\qquad 
a_2^{\rm eff}=c_2+c_1\left({1\over N_c}+\chi_2\right),
\en
where $c_{1,2}$ are the Wilson coefficients of the spectator 4-quark 
operators,
and nonfactorizable contributions are characterized by the parameters 
$\chi_1$ and $\chi_2$. Taking the decay $B^-\to D^0\pi^-$ as an example, 
we have \cite{Soares,Kamal96,Neubert}
\be
\chi_1 =\vp_8^{(BD,\pi)}+{a_1\over c_2}\vp_1^{(BD,\pi)},\qquad
\chi_2 =\vp_8^{(B\pi,D)}+{a_2\over c_1}\vp_1^{(B\pi,D)},
\en
where
\be
\vp_1^{(BD,\pi)} &=& {\la D^0\pi^-|(\bar du)_\vma(\bar cb)_\vma|B^-
\ra_{ nf}\over \la D^0\pi^-|(\bar du)_\vma(\bar cb)_\vma|B^-\ra_f}=
{\la D^0\pi^-|(\bar du)_\vma(\bar cb)_\vma|B^-
\ra\over \la \pi^-|(\bar du)_\vma|0\ra\la D^0|(\bar cb)_\vma|B^-\ra}-1,  
\non\\
\vp_8^{(BD,\pi)} &=& {1\over 2}\,{\la D^0\pi^-|(\bar d\lambda^a u)_\vma
(\bar c\lambda^a b)_\vma|B^-
\ra\over \la \pi^-|(\bar du)_\vma|0\ra\la D^0|(\bar cb)_\vma|B^-\ra},
\en
are nonfactorizable terms originated from color-singlet and color-octet
currents, respectively, $(\bar q_1q_2)_\vma\equiv \bar q_1\gamma_\mu(1-\gamma
_5)q_2$, and $(\bar q_1\lambda^a q_2)_\vma\equiv \bar q_1\lambda^a \gamma_\mu
(1-\gamma_5)q_2$. The subscript `f' and `nf' in Eq.~(1.3) stand for
factorizable and nonfactorizable contributions, respectively, and the
superscript $(BD,\pi)$ in Eq.~(1.2) means that the pion is factored out in the
factorizable amplitude of $B\to D\pi$ and likewise for the superscript
$(B\pi, D)$. In the large-$N_c$ limit, $\vp_1={\cal O}(1/N_c^2)$
and $\vp_8={\cal O}(1/N_c)$ \cite{Neubert}. Therefore, the 
nonfactorizable term $\chi$ in the $N_c\to \infty$ limit is dominated 
by color octet-octet operators. Since $|c_1/c_2|\gg 1$, it is evident from 
Eq.~(1) that even a small amount of nonfactorizable contributions will have a
significant effect on the color-suppressed class-II amplitude.
If $\chi_{1,2}$ are universal (i.e. process independent) in
charm or bottom decays, then we still have a new factorization scheme
in which the decay amplitude is expressed in terms of factorizable 
contributions multiplied by the universal effective parameters
$a_{1,2}^{\rm eff}$. (For $B\to VV$ decays, new factorization implies
that nonfactorizable terms contribute in equal weight to all partial wave
amplitudes so that $a_{1,2}^{\rm eff}$ {\it can} be defined.) The first
systematical study of nonleptonic weak decays of heavy mesons within 
the framework of the generalized factorization was carried 
out by Bauer, Stech, and Wirbel \cite{BSW87}. Phenomenological analyses
of two-body decay data of $D$ and $B$ mesons indicate that while
the generalized factorization hypothesis in general works reasonably well, 
the effective parameters $a_{1,2}^{\rm eff}$ do show some variation from
channel to channel, especially for the weak decays of charmed mesons
 \cite{Cheng94,Kamal96,Cheng96}.
An eminent feature emerged from the data analysis is that $a_2^{\rm eff}$ 
is negative in charm decay, whereas it becomes positive in the two-body decays
of the $B$ meson \cite{Cheng94,CT95,Neubert}:
\be
a_2^{\rm eff}(D\to\overline{K}\pi)\sim -0.50\,, \qquad a_2^{\rm eff}(B\to 
D\pi)\sim 0.26\,.
\en
It should be stressed that since the magnitude of $a_{1,2}$ depends on the 
model results for form factors, the above values of $a_2$ should be 
considered as representative ones.
The sign of $a_2^{\rm eff}$ is fixed by the observed destructive
interference in $D^+\to\overline{K}^0\pi^+$ and constructive interference in
$B^-\to D^0\pi^-$. Eq.~(1.4) then leads to
\be
\chi_2(\mu\sim m_c;~D\to \overline{K}\pi)\sim -0.36\,, \qquad \chi_2(\mu
\sim m_b;~B\to D\pi)\sim 0.11\,.
\en
In general the determination of $\chi_2$ is easier and more reliable 
than $\chi_1$. The observation $|\chi_2(B)|\ll|\chi_2(D)|$ is consistent
with the intuitive picture that soft gluon effects become stronger when the
final-state particles move slower, allowing more time for significant
final-state interactions after hadronization \cite{Cheng94}.

   Phenomenologically, it is often to treat the number of colors $N_c$ as
a free parameter and fit it to the data. Theoretically, this amounts to
defining an effective number of colors $\nc$, called $1/\xi$ in \cite{BSW87},
by 
\be
1/N_c^{\rm eff}\equiv (1/N_c)+\chi.
\en
It is clear from Eq.~(1.5) that
\be
N_c^{\rm eff}(D\to \overline{K}\pi) \gg 3,\qquad N_c^{\rm eff}(B\to D\pi)
\approx 2.
\en
Consequently, the empirical rule of discarding 
subleading $1/N_c$ terms formulated in the large-$N_c$ approach \cite{Buras}
is justified for exclusive charm decay; the dynamical origin of the
$1/N_c$ expansion comes from the fact that the Fierz $1/N_c$ terms 
are largely compensated by nonfactorizable effects in charm decay.
Since the large-$N_c$ approach implies $a_2^{\rm eff}\sim c_2$ and since
$a_2^{\rm eff}$ is observed to be positive in $B^-\to D^0(\pi^-,\rho^-)$ 
decays, one may wonder why is the $1/N_c$ expansion no longer 
applicable to the $B$ meson ? Contrary to the common belief,
a careful study shows this is not the case. As pointed out in \cite{Neubert}, 
the large-$N_c$ color counting rule for the Wilson
coefficient $c_2(\mu)$ is different at $\mu\sim m_b$ and $\mu\sim m_c$ due
to the presence of the large logarithm at $\mu\sim m_c$. More specifically,
$c_2(m_b)={\cal O}(1/N_c)$ and $c_2(m_c)={\cal O}(1)$. Recalling that
$c_1={\cal O}(1)$, it follows that in the large-$N_c$ limit \cite{Neubert}:
\be
a_2^{\rm eff}=\cases{ c_2(m_c)+{\cal O}(1/N_c)   & for~the~$D$~meson,  \cr
c_2(m_b)+c_1(m_b)\left({1\over N_c}+\vp_8(m_b)\right)+{\cal O}(1/N_c^3) &
for~the~$B$~meson.   \cr}
\en
Therefore, {\it a priori} the $1/N_c$ expansion does not demand a 
negative $a_2^{\rm eff}$ 
for bottom decay ! and $N_c^{\rm eff}(B\to D\pi)\sim 2$ is not in conflict
with the large-$N_c$ approach ! It should be remarked that although 
$\chi_2$ is positive in two-body decays of the $B$ meson, some theoretical 
argument suggests that it may become negative in high multiplicity 
decay modes \cite{Neubert}.

    Thus far the nonfactorizable effect is discussed at the purely
phenomenological level. It is thus important to have a theoretical 
estimate of $\chi_i$ even approximately. Unfortunately, all existing
theoretical calculations based on the QCD sum rule \cite{BS}, though confirm
the cancellation between the $1/N_c$ Fierz terms and nonfactorizable soft
gluon effects \cite{blok1}, tend to predict a negative $\chi$ in $\bar 
B^0\to D^+\pi^-,~D^0\pi^0$ and $B\to J/\psi K$ decays. This tantalizing 
issue should be
clarified and resolved in the near future. It is interesting to remark that,
relying on a different approach, namely,
the three-scale PQCD factorization theorem, to tackle the nonfactorizable 
effect, one of us and Li \cite{Li} are able to
explain the sign change of $\chi_2$ from bottom to charm decays.

   For $B$ meson decay, the effective parameters $a_{1,2}^{\rm eff}$ have
been determined so far only for $B\to D(\pi,\rho)$ and $B\to J/\psi K$ 
where nonfactorizable effects amount to having $N_c
^{\rm eff}\sim 2$. Recently, several exclusive charmless rare $B$ decay 
modes have been reported for the first time by CLEO 
\cite{Smith97,CLEO,CLEOomega1,CLEOpik,CLEOeta',Smith}
and many of them are dominated by the
penguin mechanism. It is thus important to know (i) does the constructive
interference of tree amplitudes persist in class-III charmless $B$ decay ?
(class-III transitions receive contributions from both external 
and internal $W$ emissions), and 
(ii) is $N_c^{\rm eff}$ the same in spectator and penguin amplitudes ?
In the literature it is customary to assume that $\nc$ behaves in the same 
way in
the penguin and non-penguin amplitudes. The decay rate of the rare $B$ 
decays is then studied as a function of $1/\nc$. In the present paper, we 
shall see that, theoretically and experimentally, $\nc(V+A)$ in the 
penguin amplitude induced by the $(V-A)(V+A)$ quark operators is
different from $\nc(V-A)$ in the tree or penguin amplitude induced by
the $(V-A)(V-A)$ operators.
\footnote{In \cite{CT97b} we have assumed that $\nc(V-A)
\approx\nc(V+A)\approx 2$. The present paper is an improved 
version of \cite{CT97b}.}
We find that $\nc(V+A)$ in penguin-dominated charmless $B$ decays is clearly
larger than $\nc(V-A)$ extracted from spectator-dominated processes. 
Therefore, the nonfactorizable effect in tree and penguin diagrams 
behaves in a different manner.
This observation is the key element for understanding the CLEO measurement
of $B^\pm\to\eta' K^\pm$ and $B^0\to\eta' K^0$. By treating $\nc(V-A)$
and $\nc(V+A)$ differently, the data of $B\to \eta' K$ can be explained
in the framework of the Standard Model without resorting to new mechanisms
or new physics beyond the Standard Model.

   This paper is organized as follows. in Sec.~II we sketch the starting
point of the effective Hamiltonian and emphasize that vertex and penguin
corrections to the four-quark operators should be combined together with
the Wilson coefficients to render the resulting physical amplitude 
independent of the choice of the renormalization scheme and scale. Then 
we extract the information of $\nc(V-A)$ from spectator dominated charmless
$B$ decays $B^\pm\to\omega\pi^\pm$ and $B\to\pi\pi$ in Sec.~III and 
$\nc(V+A)$ from the penguin dominated process $B\to\phi K$
in Sec.~IV. In Sec.~V we demonstrate that the measurement of $B\to\eta' K$ 
also favors a different treatment of $\nc(V-A)$ and $\nc(V+A)$. In Sec.~VI
we point out a difficulty with $B^\pm\to\omega K^\pm$ within the present
framework. Conclusions and discussions are presented in Sec.~VII.

\section{Calculational Framework}

   We briefly sketch in this section the calculational framework.
The relevant effective $\Delta B=1$ weak Hamiltonian is
\be
{\cal H}_{\rm eff}(\Delta B=1) = {G_F\over\sqrt{2}}\Big[ V_{ub}V_{uq}^*(c_1
O_1^u+c_2O_2^u)+V_{cb}V_{cq}^*(c_1O_1^c+c_2O_2^c)
-V_{tb}V_{tq}^*\sum^{10}_{i=3}c_iO_i\Big]+{\rm h.c.},
\en
where $q=d,s$, and
\be
&& O_1^u= (\bar ub)_\vma(\bar qu)_\vma, \qquad\qquad\qquad\qquad~~
O_2^u = (\bar qb)_\vma(\bar uu)_\vma, \non \\
&& O_{3(5)}=(\bar qb)_\vma\sum_{q'}(\bar q'q')_{\vma(\vpa)}, \qquad  \qquad
O_{4(6)}=(\bar q_\alpha b_\beta)_\vma\sum_{q'}(\bar q'_\beta q'_\alpha)_{
\vma(\vpa)},   \\
&& O_{7(9)}={3\over 2}(\bar qb)_\vma\sum_{q'}e_{q'}(\bar q'q')_{\vpa(\vma)},
  \qquad O_{8(10)}={3\over 2}(\bar q_\alpha b_\beta)_\vma\sum_{q'}e_{q'}(\bar 
q'_\beta q'_\alpha)_{\vpa(\vma)},   \non
\en   
with $O_3$-$O_6$ being the QCD penguin operators and $O_{7}$-$O_{10}$ 
the electroweak penguin operators.
As noted in passing, in order to ensure the renormalization-scale and -scheme
independence for the physical amplitude, the matrix elements of 4-quark
operators have to be evaluated in the same renormalization scheme as
that for Wilson coefficients and renormalized at the same scale $\mu$. 

In full theory, the leading QCD correction to the weak transition is of the 
form $\alpha_s\ln(M_W^2/\!-p^2)$ for massless quarks, where $p$ is the 
off-shell momentum of external quark lines and its magnitude $-p^2$ depends 
on the system under consideration. For example, $-p^2\sim m_b^2$ in the 
energetic two-body charmless $B$ decays.
The merit of the effective Hamiltonian approach is that one 
can choose a renormalization scale $\mu$ so that the leading logarithmic
correction $\ln(M^2_W/\!-p^2)=\ln(M^2_W/\mu^2)+\ln(\mu^2/\!-p^2)$ is 
decomposed 
in such a way that the large logarithmic term $\ln(M^2_M/\mu^2)$ is
lumped into the Wilson coefficient function $c(\mu)$ and summed over 
to all orders in $\alpha_s$ using 
the renormalization group equation, while the  logarithmic correction 
$\ln(\mu^2/\!-p^2)$ to the matrix element $\la O(\mu)\ra$ is small (for
a review, see \cite{Buras96}). Since $O(\mu)$ is the 
four-quark operator renormalized at the
scale $\mu$, its hadronic matrix element is related to the tree level one 
via
\be
\la O(\mu)\ra=\,g(\mu)\la O\ra_{\rm tree},
\en
with
\be
g(\mu)\sim 1+\alpha_s(\mu)\left(\gamma\ln{\mu^2\over -p^2}+c\right)
\en
for current-current operators, where we have included the non-logarithmic 
constant contribution
$c$ since the logarithmic contribution $\ln(\mu^2/\!-p^2)$ is small when 
$\mu^2\sim -p^2$ and hence the momentum-independent constant term cannot
be neglected. It follows that schematically
\be
\la {\cal H}_{\rm eff}\ra=\,c(\mu)g(\mu)\la O\ra_{\rm tree}=c^{\rm eff}\la 
O\ra_{\rm tree}.
\en
To the next-to-leading order (NLO), $c(\mu)$ depends on the renormalization 
scheme chosen, so does the constant $c$ in $g(\mu)$. However,
the effective Wilson coefficient $c^{\rm eff}$ is independent of 
the choice of the renormalization scheme and scale.  It should be stressed 
that, except for the lattice QCD, model calculations of 
the hadronic matrix elements are actually performed for $\la O\ra_{\rm tree}$ 
rather than for $\la O(\mu)\ra$. (Quark model calculation of
$\la O\ra_{\rm tree}$, for example, may involve an implicit low energy 
scale, but it has nothing to do with the renormalization scale $\mu$.)
For example, in the factorization approximation, the 
matrix element $\la O\ra_{\rm fact}$ is scale independent and hence it cannot 
be identified with $\la O(\mu)\ra$. Therefore, it is
important to evaluate $g(\mu)$, the perturbative corrections
to the four-quark operators at the scale $\mu$.

As emphasized above, before applying factorization or carrying out 
any model calculation of hadronic matrix elements,
it is necessary to incorporate QCD
and electroweak corrections to the operators:
\be
\la O_i(\mu)\ra=\left[\,{\rm I}+{\alpha_s(\mu)\over 4\pi}\m_s(\mu)+{\alpha
\over 4\pi}\m_e(\mu)\right]_{ij}\la O_j\ra_{\rm tree},
\en
so that $c_i(\mu)\la O_i(\mu)\ra= {c}_i^{\rm eff}\la O_i\ra_{\rm tree}$, where
\be
{c}_i^{\rm eff}=\left[\,{\rm I}+{\alpha_s(\mu)\over 4\pi}\m_s^T(\mu)+{\alpha
\over 4\pi}\m_e^T(\mu)\right]_{ij}c_j(\mu).
\en
Then the factorization approximation is applied to the hadronic matrix
elements of the operator $O$ at tree level. Perturbative QCD and 
electroweak corrections to the matrices $\hat m_s$ and $\hat m_e$ 
from vertex diagrams and penguin diagrams have been
calculated in \cite{Buras92,Flei,Kramer,Ali}. It should be remarked that
although the penguin coefficients $c_{3}-c_{10}$ are governed by the penguin
diagrams with $t$ quark exchange, the effective Wilson coefficients do 
incorporate the effects of the penguin diagrams with internal
$u$ and $c$ quarks induced by the current-current operator $O_1$.
For example \cite{Ali},
\be
c_6^{\rm eff}=\,c_6(\mu)-{\alpha_s(\mu)\over 8\pi}\left[ {\lambda_c\over
\lambda_t}\tilde{G}(m_c,k,\mu)+{\lambda_u\over \lambda_t}\tilde{G}(m_u,k,\mu)
\right]c_1(\mu)+\cdots,
\en
where $\lambda_i=V_{ib}V^*_{iq}~(q=d,s)$, $\tilde{G}(m_q,k,\mu)={2\over 3}
\kappa-G(m_q,k,\mu)$, $\kappa$ is a constant depending on the 
renormalization scheme, $k$ is the gluon's virtual momentum, and
\be
G(m,k,\mu)=-4\int^1_0dxx(1-x)\ln\left({m^2-k^2x(1-x)\over \mu^2}\right).
\en
For $b\to s$ transitions, $|\lambda_u|\ll|\lambda_t|$, $\lambda_c\sim 
-\lambda_t$, and hence
\be
c_6^{\rm eff}=c_6(\mu)+{\alpha_s(\mu)\over 8\pi}\tilde{G}(m_c,k,\mu)c_1(\mu)
+\cdots.
\en
The importance of the so-called ``charming'' penguins for $b\to s$ transition 
was emphasized recently (and probably over-emphasized) in \cite{Ciuchini}.

Using the next-to-leading order $\Delta B=1$ Wilson 
coefficients obtained in the 't Hooft-Veltman (HV) scheme and the naive 
dimension regularization (NDR) scheme at $\mu={m}_b(m_b)$, 
$\Lambda^{(5)}_{\overline{\rm MS}}=225$ 
MeV and $m_t=170$ GeV in Table 22 of \cite{Buras96}, we obtain the effective
renormalization-scheme and -scale independent Wilson coefficients 
$c_i^{\rm eff}$ at $k^2=m_b^2/2$: 
\footnote{We use the complete expressions of $\hat{m}_s(\mu)$ given in
\cite{Ali} and $\hat{m}_e(\mu)$ in \cite{Flei} to evaluate
$c^{\rm eff}_i$. Note that while our $c_{1-6}^{\rm eff}$ are consistent
with the numerical results given in \cite{Ali,Deshpande1}, our values for
$c^{\rm eff}_{7-10}$ are  different from that shown in
\cite{Deshpande1}.}
\be
&& {c}_1^{\rm eff}=1.149, \qquad\qquad\qquad\qquad\qquad {c}_2^{\rm eff}=
-0.325,   \non \\
&& {c}_3^{\rm eff}=0.0211+i0.0045, \qquad\qquad\quad\,{c}_4^{\rm eff}=-0.0450
-i0.0136, \non  \\
&& {c}_5^{\rm eff}=0.0134+i0.0045, \qquad\qquad\quad\,{c}_6^{\rm eff}=-0.0560
-i0.0136, \non  \\
&& {c}_7^{\rm eff}=-(0.0276+i0.0369)\alpha, \qquad \quad {c}_8^{\rm eff}=
0.054\,\alpha, \non  \\
&& {c}_9^{\rm eff}=-(1.318+i0.0369)\alpha, \qquad \quad ~\,{c}_{10}^{\rm eff}=
0.263\,\alpha.
\en
  Two important remarks are in order. First of all, $c_{1,2}^{\rm eff}$ 
are surprisingly very close to the leading order Wilson coefficients: 
$c_1^{\rm LO}=1.144$ and $c_2^{\rm LO}=-0.308$ at $\mu={m}_b(m_b)$ 
\cite{Buras96}, recalling that $c_2^{\rm NDR}=-0.185$ and 
$c_2^{\rm HV}=-0.228$ at NLO \cite{Buras96} deviate substantially from the 
leading order values. This means that 
$\la O_{1,2}(\mu)\ra\approx \la O_{1,2}\ra_{\rm tree}$. Hence, it 
explains why the conventional way of applying the Wilson coefficients at
leading order and evaluating the matrix elements of current-current operators 
at tree level is ``accidentally'' justified provided that $\mu^2\sim -p^2$.
Second, comparing (2.11) with the 
leading-order penguin coefficients \cite{Buras96}
\be
c_3^{\rm LO}=0.014, \qquad c_4^{\rm LO}=-0.030, \qquad c_5^{\rm LO}=0.009,
\qquad c_6^{\rm LO}=-0.038
\en
at $\mu={m}_b(m_b)$, 
we see that Re$(c^{\rm eff}_{3-6})\approx {3\over 2} c_{3-6}^{\rm LO}(\mu)$. 
This implies that, contrary to the case of current-current operators,
penguin corrections to the current-current operators give important
contributions to the QCD penguin operators.  
This means that the decay rates of charmless $B$ decay modes
dominated by penguin diagrams will be too small by a factor of 
$\sim (1.5)^2=2.3$ if 
only leading-order penguin coefficients are employed for calculation.

    We shall see later that running quark masses appear in the matrix
elements of $(S-P)(S+P)$ penguin operators through the use of
equations of motion. The running quark mass should be applied at the 
scale $\mu\sim m_b$ because the energy release in the
energetic two-body charmless decays of the $B$ meson is of order $m_b$.
Explicitly, we use \cite{Fusaoku}
\be
&& m_u(m_b)=3.2\,{\rm MeV},  \qquad m_d(m_b)=6.4\,{\rm MeV},  \qquad
m_s(m_b)=105\,{\rm MeV},  \non \\
&&  m_c(m_b)=0.95\,{\rm GeV},  \qquad
m_b(m_b)=4.34\,{\rm GeV},  
\en
in ensuing calculation, where we have applied $m_s=150$ MeV at $\mu=1$ GeV.

   It is convenient to parametrize the quark mixing matrix in terms of the
Wolfenstein parameters: $A,~\lambda,~\rho$ and $\eta$ \cite{Wolf}, where
$A=0.804$ and $\lambda=0.22$. In the present paper we employ two 
representative values for $\rho$ and $\eta$: (i) $\rho=0.16,~\eta=0.34$,
and (ii) $\rho=-0.12,~\eta=0.35$. Both of them satisfy the constraint
$\sqrt{\rho^2+\eta^2}=0.37$. A recent analysis of all available experimental 
constraints imposed on the Wolfenstein parameters yields \cite{Parodi}
\be
\bar{\rho}=\,0.156\pm 0.090\,,   \qquad \bar{\eta}=\,0.328\pm 0.054,
\en
where $\bar{\rho}=\rho(1-{\lambda^2\over 2})$ and $\bar\eta=\eta(1-{\lambda
^2\over 2})$, and it implies that the negative $\rho$ region is excluded at 
93\% C.L..

\section{Nonfactorizable effects in spectator amplitudes}
  The combinations of the effective Wilson coefficients $a_{2i}=
{c}_{2i}^{\rm eff}+{1\over N_c}{c}_{2i-1}^{\rm eff},~a_{2i-1}=
{c}_{2i-1}^{\rm eff}+{1\over N_c}{c}^{\rm eff}_{2i}$ $(i=1,\cdots,5)$ 
appear in the decay amplitudes. As discussed in the Introduction, 
nonfactorizable effects in the 
decay amplitudes of $B\to PP,~VP$ can be absorbed into the parameters
$a_i^{\rm eff}$. This amounts to replacing $N_c$ in $a_i$
by $(N_c^{\rm eff})_i$. (It must be
emphasized that the factor of $N_c$ appearing in any place other than 
$a_i$ should {\it not} be replaced by $\nc$.) Explicitly,
\be
a_{2i}^{\rm eff}={c}_{2i}^{\rm eff}+{1\over (N_c^{\rm eff})_{2i}}{c}_{2i-1}^{
\rm eff}, \qquad \quad a_{2i-1}^{\rm eff}=
{c}_{2i-1}^{\rm eff}+{1\over (N_c^{\rm eff})_{2i-1}}{c}^{\rm eff}_{2i}.
\en
It is customary to assume in the literature that $(N_c^{\rm eff})_1
\approx (N_c^{\rm eff})_2\cdots\approx (N_c^{\rm eff})_{10}$
so that the subscript $i$ can be dropped. A closer investigation shows 
that this is not the case.  Consider an operator of the form $O=\bar{q}_1
^\alpha\Gamma q_2^\beta\,\bar{q}_3^\beta\Gamma' q_4^\alpha$ which arises 
from the Fierz transformation of a singlet-singlet operator with $\Gamma$ and
$\Gamma'$ being some combinations of Dirac matrices. Applying the identity
\be
O=\,{1\over 3}\bar{q}_1\Gamma q_2\,\bar{q}_3\Gamma' q_4+{1\over 2}\bar{q}_1
\lambda^a\Gamma q_2\,\bar{q}_3\lambda^a\Gamma' q_4,
\en
to the matrix element of $M\to P_1P_2$ leads to (assuming the quark content
$\bar q_1 q_2$ for $P_1$)
\be
\la P_1P_2|O|M\ra &=& {1\over 3}\la P_1|\bar q_1\Gamma q_2|0\ra\la P_2|
\bar q_3\Gamma' q_4|M\ra+{1\over 3}\la P_1P_2|\bar q_1\Gamma q_2\,
\bar q_3\Gamma' q_4|M\ra_{nf}   \non\\
&+& {1\over 2}\la P_1P_2|\bar q_1\lambda^a\Gamma q_2\,\bar q_3\lambda^a\Gamma'
q_4|M\ra.
\en
The nonfactorizable effects due to octet-octet and singlet-singlet operators
are characterized by the parameters $\vp_8$ and $\vp_1$, respectively,
as shown in Eq. (1.3):

\be
\vp_1 &=& {\la P_1P_2|\bar q_1\Gamma q_2\,\bar q_3\Gamma' 
q_4 |M\ra_{ nf}\over \la P_1P_2|\bar q_1\Gamma q_2\,\bar q_3\Gamma' 
q_4 |M\ra_f}\times {\la P_1P_2|\bar q_1\Gamma q_2\,\bar q_3\Gamma' 
q_4 |M\ra_f\over \la P_1P_2|(\bar q_1q_2)_\vma(\bar q_3 
q_4)_\vma |M\ra_f},   \non \\
\vp_8 &=& {1\over 2}\,{\la P_1P_2|\bar q_1\lambda^a \Gamma q_2\,\bar q_3
\lambda^a\Gamma' q_4|M\ra\over \la P_1P_2|\bar q_1\Gamma q_2\,\bar q_3\Gamma' 
q_4|M\ra_f }\times {\la P_1P_2|\bar q_1\Gamma q_2\,\bar q_3\Gamma' 
q_4 |M\ra_f\over \la P_1P_2|(\bar q_1q_2)_\vma(\bar q_3 
q_4)_\vma |M\ra_f}\,.
\en
However, the Fierz transformation of the $(V-A)(V+A)$ operators $O_{5,6,7,8}$
is quite different from that of $(V-A)(V-A)$ operators $O_{1,2,3,4}$
and $O_{9,10}$; that is,
\be
(V-A)(V+A) &\to& -2(S-P)(S+P), \non \\
(V-A)(V-A) &\to& (V-A)(V-A).
\en
Therefore, $\Gamma$ and $\Gamma'$ are the combinations of the Dirac matrices
1 and $\gamma_5$ for the Fierz transformation of $(V-A)(V+A)$ 
operators, and the combinations of $\gamma_\mu$ and $\gamma_\mu\gamma_5$ for
$(V-A)(V-A)$ operators. As a result, nonfactorizable effects in the matrix
elements of $(V-A)(V+A)$ operators are {\it a priori} different from that of 
$(V-A)(V-A)$ operators, i.e. $\chi(V+A)\neq \chi(V-A)$. Since $1/N_c^{\rm 
eff}=1/N_c+\chi$ [cf. Eq.~(1.6)], theoretically it is expected that
\be
&& N_c^{\rm eff}(V-A)\equiv
\left(N_c^{\rm eff}\right)_1\approx\left(N_c^{\rm eff}\right)_2\approx
\left(N_c^{\rm eff}\right)_3\approx\left(N_c^{\rm eff}\right)_4\approx
\left(N_c^{\rm eff}\right)_9\approx
\left(N_c^{\rm eff}\right)_{10},   \non\\
&& N_c^{\rm eff}(V+A)\equiv
\left(N_c^{\rm eff}\right)_5\approx\left(N_c^{\rm eff}\right)_6\approx
\left(N_c^{\rm eff}\right)_7\approx
\left(N_c^{\rm eff}\right)_8,
\en
and $N_c^{\rm eff}(V+A)\neq N_c^{\rm eff}(V-A)$ in general. In principle, 
$N_c^{\rm eff}$ can vary from channel to channel, as in the case of charm
decay. However, in the energetic two-body $B$ decays, $\nc$
is expected to be process insensitive as supported by data 
\cite{Neubert}. As stressed in the Introduction, if $\nc$ is process 
independent, then we have a generalized factorization. Contrary to the
naive one, the improved factorization does incorporate nonfactorizable
effects in a process independent form. For example, $\chi_1=\chi_2=-{1
\over 3}$ in the large-$N_c$ approximation of factorization.

   The unknown parameters $(\nc)_i$ in charmless $B$ decays in principle can 
be determined if the decay rates are measured for a handful of decay modes 
with sufficient
accuracy. Due to the limited data and limited significance available at 
present we shall use (3.6) 
and the experimental result for $\nc(B\to D\pi)$ as a guidance to 
determine $(\nc)_i$. To begin with, we focus
in this section the decay modes dominated by the spectator diagrams induced
by the current-current operators $O_1$ and $O_2$. In particular, we 
would like to study these modes which are sensitive to the interference
between external and internal $W$-emission amplitudes. The fact that
$\nc<3.5$ ($\nc>3.5$) implies a positive (negative) $a_2^{\rm eff}$ and 
hence a constructive (destructive) interference 
will enable us to differentiate between them. Good
examples are the class-III modes: $B^\pm\to \omega\pi^\pm,~\pi^0\pi^\pm,~\eta
\pi^\pm,~\pi^0\rho^\pm,\cdots$, etc.

   We first consider the decay $B^-\to\omega\pi^-$. Under the generalized 
factorization, its decay amplitude is given by
\be
A(B^-\to\omega\pi^-) &=& {G_F\over\sqrt{2}}\Bigg\{ V_{ub}V_{ud}^*\left(a_1
X^{(B\omega,\pi)}+a_2X^{(B\pi,\omega)}+2a_1X^{(B,\pi\omega)}\right)  \non \\
&-& V_{tb}V_{td}^*\Bigg[ \left(a_4+a_{10}-2(a_6+a_8){m_\pi^2\over (m_b+m_u)
(m_u+m_d)}\right)X^{(B\omega,\pi)}   \non \\
&+& {1\over 2}\left(4a_3+2a_4+4a_5+a_7+a_9-a_{10}\right)X^{(B\pi,\omega)}  
\non \\
&+& 2\left(a_4+a_{10}-2(a_6+a_8){m_B^2\over (m_b+m_u)
(m_u+m_d)}\right)X^{(B,\pi\omega)} \Bigg]\Bigg\},
\en
where we have dropped the superscript ``eff'' for convenience, and
the notation $X^{(B\omega,\pi)}$, for example, denotes the factorization
amplitude with the $\pi$ meson being factored out:
\be
X^{(B\omega,\pi)} &\equiv& \la \pi^-|(\bar du)_\vma|0\ra\la\omega|(\bar ub)_
\vma|B^-\ra,    \non \\
X^{(B\pi,\omega)} &\equiv& \la \omega|(\bar uu)_\vma|0\ra\la \pi^-|(\bar db)_
\vma|B^-\ra,   \non \\
X^{(B,\pi\omega)} &\equiv& \la\pi^-\omega |(\bar du)_\vma|0\ra\la 0|(\bar ub)_
\vma|B^-\ra.
\en
Note that in the penguin amplitude, the term $X^{(B,\pi\omega)}$ arises from
the space-like penguin diagram.
Using the following parametrization for decay constants and form factors:
\footnote{Once the one-body matrix elements are defined, one can apply heavy
quark symmetry to the two-body matrix elements for heavy-to-heavy 
transition to show that all the form factors defined in (3.9) are positively
defined at $q^2\geq 0$ and that the relative signs between two-body and 
one-body matrix elements are fixed. In this way, we find that  
the vector form factor $V(q^2)$ defined by Bauer, Stech and Wirbel 
\cite{BSW85} has a sign opposite to ours. Note that our convention is
$\epsilon_{0123}=1$.}
\be
\la 0|A_\mu|P(q)\ra &=& if_Pq_\mu, \qquad \la 0|V_\mu|V(p,\vp)\ra=f_Vm_V\vp_
\mu,   \non \\
\la P'(p')|V_\mu|P(p)\ra &=& \left(p_\mu+p'_\mu-{m_P^2-m_{P'}^2\over q^2}\,q_
\mu\right) F_1(q^2)+F_0(q^2)\,{m_P^2-m_{P'}^2\over q^2}q_\mu,   \non \\
\la V(p',\vp)|V_\mu|P(p)\ra &=& {2\over m_P+m_V}\,\epsilon_{\mu\nu\alpha
\beta}\vp^{*\nu}p^\alpha p'^\beta V(q^2),   \non \\
\la V(p',\vp)|A_\mu|P(p)\ra &=& i\Big[ (m_P+m_V)\vp_\mu A_1(q^2)-{\vp\cdot
p\over m_P+m_V}\,(p+p')_\mu A_2(q^2)    \non \\
&& -2m_V\,{\vp\cdot p\over q^2}\,q_\mu\big[A_3(q^2)-A_0(q^2)\big]\Big],
\en
where $q=p-p'$, $F_1(0)=F_0(0)$, $A_3(0)=A_0(0)$, and
\be
A_3(q^2)=\,{m_P+m_V\over 2m_V}\,A_1(q^2)-{m_P-m_V\over 2m_V}\,A_2(q^2),
\en
we obtain
\be
X^{(B\omega,\pi)} &=& -2f_\pi m_\omega A_0^{B\omega}(m_\pi^2)\vp\cdot p_B, 
\non\\
X^{(B\pi,\omega)} &=& -\sqrt{2}f_\omega m_\omega F_1^{B\pi}(m_\omega^2)
\vp\cdot p_B. 
\en

For the $q^2$ dependence of form factors in the region where $q^2$ is not 
too large, we shall use the pole dominance ansatz, namely,
\be
f(q^2)=\,{f(0)\over \left(1-{q^2/m^2_*}\right)^n},
\en
where $m_*$ is the pole mass given in \cite{BSW87}. 
A direct calculation of $B\to P$ and $B\to V$ 
form factors at time-like momentum transfer is available in the relativistic 
light-front quark model \cite{CCH} with the results that
the $q^2$ dependence of the form factors $A_0,~F_1$ is a dipole behavior 
(i.e. $n=2$), while $F_0$ exhibits a monopole dependence ($n=1$). The decay 
rate is then given by
\be
\Gamma(B^-\to\pi^-\omega)={p_c\over 8\pi m_B^2}\left({(m_B^2-m^2_\pi
-m^2_\omega)^2\over 4m_\omega^2}-m_\pi^2\right)\left|{A(B^-\to\pi^-
\omega)\over \vp\cdot p_B}\right|^2,
\en
where $p_c$ is the c.m. momentum
\be
p_c=\,{\sqrt{[m_B^2-(m_\omega+m_\pi)^2][m_B^2-(m_\omega-m_\pi)^2]}\over
2m_B}\,.
\en

Since
\be
V_{ub}V_{ud}^*=A\lambda^3(\rho-i\eta), \quad V_{cb}V_{cd}^*=-A\lambda^3,
\quad V_{tb}V_{td}^*=A\lambda^3(1-\rho+i\eta),
\en
in terms of the Wolfenstein parametrization \cite{Wolf}, are of the same 
order of magnitude, it is clear that $B^-\to \omega\pi^-$ is dominated by  
external and internal $W$ emissions and that penguin contributions are
suppressed by the smallness of penguin coefficients. 
Neglecting the $W$-annihilation contribution denoted by $X^{(B,\pi\omega)}$, 
and using $f_\pi=132$ MeV, $f_\omega=195$ MeV for decay constants, 
$A_0^{B\omega}(0)=0.28/\sqrt{2},~F_1^{B \pi}(0)=0.33$ for form factors 
\cite{BSW85}, and $\tau(B^\pm)=(1.67\pm 0.04)$ 
ps \cite{LEP} for the charged $B$ lifetime, the branching ratio of 
$B^\pm\to\pi^\pm\omega$ averaged over CP-conjugate modes is shown
in Fig.~1 where we have set $\nc(V+A)=\nc(V-A)=\nc$ 
and plotted the branching ratio as a function of $1/\nc$. 
We see that the branching ratio is sensitive to $1/\nc$ and has the 
lowest value of order $2\times 10^{-6}$ at $\nc=\infty$ and then increases 
with $1/\nc$. Since experimentally \cite{CLEOomega1}
\footnote{The significance of $B^\pm\to\omega\pi^\pm$ is reduced in the
recent CLEO analysis and only an upper limit is quoted 
\cite{CLEOomega2,Smith}:
${\cal B}(B^\pm\to\pi^\pm\omega)<2.3\times 10^{-5}$.
Since ${\cal B}(B^\pm\to K^\pm\omega)=(1.5^{+0.7}_{-0.6}\pm 0.2)
\times 10^{-5}$ and ${\cal B}(B^\pm\to h^\pm\omega)=(2.5^{+0.8}_{-0.7}\pm 0.3)
\times 10^{-5}$ with $h=\pi,~K$, the central value of ${\cal B}(B^\pm\to
\pi^\pm\omega)$ remains about the same as (3.16).}
\be
{\cal B}(B^\pm\to\omega\pi^\pm)=\left(1.1^{+0.6}_{-0.5}\pm 0.2\right)
\times 10^{-5},
\en
it is evident that $1/\nc>0.35$ is preferred by the data. Because this decay 
is dominated by tree amplitudes, this in turn implies that 
\be
\nc(V-A)<2.9~~~{\rm from}~B^\pm\to\pi^\pm\omega.
\en
With the value of $\nc(V-A)$ being fixed to be 2, the branching ratio of 
$B^\pm\to\pi^\pm
\omega$ is plotted in Fig.~2 as a function of $\nc(V+A)$. We see that 
for positive $\rho$, which is preferred by the current analysis 
\cite{Parodi}, the branching ratio is of order $(0.9-1.0)\times 10^{-5}$, 
very close to the central value of the measured one.

  The fact that $\nc(V-A)<2.9$ in charmless two-body decays of the $B$ meson 
is consistent with the nonfactorizable term extracted from $B\to (D,D^*) 
\pi,~D\rho$ decays, namely $\chi\sim 0.10$ or $\nc(B\to D\pi)\approx 2$. 
Since the
energy release in the energetic two-body decays $B\to\omega\pi$, $B\to D\pi$
is of the same order of magnitude, it is thus expected that $\nc(V-A)|_{B
\to\omega\pi}\approx 2$.

The main 
uncertainty of the above analysis is the negligence of the space-like
penguins and $W$-annihilations. It is 
common to argue that $W$-annihilation is negligible due to helicity 
suppression, corresponding to form factor suppression at large momentum 
transfer, $q^2=m_B^2$ (for a recent study, see \cite{Xing}).
However, we see from Eq.~(3.7) that the space-like penguin contribution 
gains a large enhancement by a factor 
of $m_B^2/[m_b(m_u+m_d)]\approx 670$.
Therefore, there is no good reason to ignore the space-like
penguin effect \cite{Chau1} that has been largely overlooked in the 
literature. Unfortunately, we do not 
have a reliable method for estimating $W$-annihilation and hence
space-like penguins.

   We next come to the decay $B^-\to\pi^-\pi^0$ which is quite clean and 
unique in the sense that this is the only two-body charmless $B$ decay 
mode that does not receive any contributions from the QCD penguin operators.
Under the generalized factorization approximation,
\be
A(B^-\to\pi^-\pi^0)=\,{G_F\over\sqrt{2}}V_{ub}V^*_{ud}(a_1+a_2)if_\pi(m_B^2
-m_\pi^2)F_0^{B\pi^0}(m_\pi^2),
\en
with $F_0^{B\pi^0}=F_0^{B\pi^\pm}/\sqrt{2}$, where we have neglected the very
small electroweak penguin contributions. The decay rate is
\be
\Gamma(B^-\to\pi^-\pi^0)=\,{p_c\over 8\pi m_B^2}\,|A(B^-\to\pi^-\pi^0)|^2.
\en
Just like the decay $B^-\to\pi^-\omega$,
the branching ratio of $B^-\to\pi^-\pi^0$ also increases with $1/\nc$ as shown
in Fig.~3. The CLEO measurement is \cite{CLEOpik}
\be
{\cal B}(B^\pm\to\pi^\pm\pi^0)=\left(0.9^{+0.6}_{-0.5}\right)\times
10^{-5}~~<2.0\times 10^{-5}.
\en
However, the errors are so large that it is meaningless to put a 
sensible constraint 
on $\nc(V-A)$. Nevertheless, we see that in the range $0\leq 1/\nc\leq
0.5$ \cite{Ali}, $\nc(V-A)\approx 2$ is favored.

 In analogue to the decays $B\to D^{(*)}\pi(\rho)$, the interference effect of
spectator amplitudes in class-III charmless $B$ decay can be tested 
by measuring the ratios:
\be
R_1\equiv 2\,{{\cal B}(B^-\to\pi^-\pi^0)\over {\cal B}(\bar B^0\to \pi^-\pi^+
)},\qquad R_2\equiv 2\,{{\cal B}(B^-\to\rho^-\pi^0)\over {\cal B}(\bar 
B^0\to \rho^-\pi^+)},\qquad R_3\equiv 2\,{{\cal B}(B^-\to\pi^-\rho^0)\over 
{\cal B}(\bar B^0\to \pi^-\rho^+)}.
\en
Since penguin contributions are very small as we have checked numerically, 
to a good approximation we have
\be
R_1 &=& {\tau(B^-)\over\tau(B^0_d)}\left(1+{a_2\over a_1}\right)^2,  \non\\
R_2 &=& {\tau(B^-)\over\tau(B^0_d)}\left(1+{f_\pi\over f_\rho}\,{A_0^{B\rho}
(m^2_\pi)\over F_1^{B\pi}(m^2_\rho)}\,{a_2\over a_1}\right)^2,  \non\\
R_3 &=& {\tau(B^-)\over\tau(B^0_d)}\left(1+{f_\rho\over f_\pi}\,{F_1^{B\pi}
(m^2_\rho)\over A_0^{B\rho}(m^2_\pi)}\,{a_2\over a_1}\right)^2.
\en
Evidently, the ratios $R_i$ are greater (less) than unity when the 
interference is constructive (destructive). Numerically we find
\be
R_1=\cases{1.74,  \cr 0.58, \cr} \qquad R_2=\cases{1.40, \cr 0.80, \cr} \qquad
R_3=\cases{2.50 & for~$\nc=2$,  \cr  0.26 & for~$\nc=\infty$, \cr}
\en
where use of $\tau(B^0_d)=(1.57\pm 0.04)$ ps \cite{LEP}, $f_\rho=216$ MeV,
$A_0^{B\rho}(0)=0.28$ \cite{CCH} has been made. Hence, a measurement of
$R_i$ (in particular $R_3$), which has the advantage of being independent of
the parameters $\rho$ and $\eta$, will constitute a very useful test on the
effective number of colors $\nc(V-A)$. The present experimental 
information on $\overline{B}^0\to\pi^+\pi^-$ is \cite{CLEOpik}
\be
{\cal B}(\overline{B}^0\to\pi^\pm\pi^\mp)=(0.7\pm 0.4)\times 10^{-5}~~<1.5
\times 10^{-5}.
\en
As far as the experimental central value of $R_1$ is concerned, it appears 
that $1/\nc\sim
0.5$ is more favored than any other small values of $1/\nc$.

   In short, using the central values of the branching ratios for class-III
decay modes: $B\to\pi\omega,~B\to\pi\pi$, we find that within the range
$0\leq 1/\nc\leq 0.5$, $\nc(V-A)\sim 2$ is certainly more preferred. 
Measurements of class-III decays are urgently needed in order
to pin down the nonfactorizable effect in tree amplitudes.
In particular, measurements of the interference effects in charged $B$
decays $B^-\to\pi^-(\rho^-)\pi^0(\rho^0)$ will be very helpful in
determining $\nc(V-A)$.

\section{Nonfactorizable effects in penguin amplitudes}
In Sec.~III we have shown that for spectator-diagram amplitudes $\nc(V-A)
\sim 2$ is preferred, as expected. However, the nonfactorizable effect in
the penguin amplitude is not necessarily the same as that in the tree
amplitude since the chiral structure and the Fierz
transformation of the $(V-A)(V+A)$ 4-quark operators
$O_{5,6,7,8}$ are different from that
of $(V-A)(V-A)$ operators; that is, 
$\nc(V+A)$ is {\it  a priori} not the same as $\nc(V-A)$. By
studying the penguin-dominated decays $B\to\phi K$ and $B\to\phi K^*$,
we shall see that $\nc(V+A)\sim 2 $ is ruled out by the current
bound on $B^-\to K^-\phi$.

   The decay $B^-\to K^-\phi$ receives contributions from $W$-annihilation and
penguin diagrams:
\be
A(B^-\to K^-\phi) &=& {G_F\over\sqrt{2}}\Bigg\{ V_{ub}V_{us}^*a_1X^{(B,K\phi)}
- V_{tb}V_{ts}^*\Bigg[\left( a_3+a_4+a_5-{1\over 2}(a_7+a_9+a_{10})\right)
X^{(BK,\phi)}    \non \\
&+& \left( a_4+a_{10}-2(a_6+a_8){m_B^2\over (m_b+m_u)
(m_s+m_u)}\right)X^{(B,K\phi )}   \Bigg]\Bigg\},
\en
where
\be
X^{(BK,\phi)} &\equiv& \la \phi|(\bar ss)_\vma|0\ra\la K^-|(\bar sb)_\vma|B^-
\ra=-2f_\phi m_\phi F_1^{BK}(m_{\phi}^2)(\vp\cdot p_B),   \non \\
X^{(B,K\phi)} &\equiv& \la\phi K^-|(\bar su)_\vma|0\ra\la 0|(\bar ub)_\vma|
B^-\ra.
\en
Neglecting $W$-annihilation and space-like penguin diagrams and using
$f_\phi=237$ MeV, $F_1^{BK}(0)=0.34$ \cite{CCH}, we plot in 
Fig.~4 the branching ratio of $B^\pm\to\phi K^\pm$ against 
$1/\nc$ for two different cases: the dotted curve for the free parameter
$\nc(V+A)=\nc(V-A)=\nc$ and the solid curve with $\nc(V-A)$ being fixed at
the value of 2. In either case, it is clear that $\nc(V+A)=2$
is evidently excluded from the present CLEO upper limit \cite{CLEOomega2}
\be
{\cal B}(B^\pm\to\phi K^\pm)< 0.5\times 10^{-5}.
\en
A similar observation was also made in \cite{Deshpande2}. The conclusion
that $\nc(V+A)\neq 2$ will be further reinforced if the decay rate of 
$B^\pm\to
\phi K^\pm$ is enhanced by the space-like penguins. From Fig.~4 we also
see that $1/\nc(V+A)<0.23$ or $\nc(V+A)>4.3\,$. Note that this 
constraint is subject to the corrections from space-like penguin and
$W$-annihilation contributions. At any rate, it is safe to conclude that
\be
\nc(V+A)>\nc(V-A).
\en

  The branching ratio of $B\to\phi K^*$, the average of $\phi K^{*-}$ and
$\phi K^{*0}$ modes, is also measured recently by CLEO with the result
\cite{CLEOomega2}
\be
{\cal B}(B\to\phi K^*)\equiv {1\over 2}\left[{\cal B}(B^\pm\to\phi K^{*\pm})
+{\cal B}(B^0\to\phi K^{*0})\right]
=\left(1.1^{+0.6}_{-0.5}\pm 0.2\right)\times 10^{-5}.
\en
As emphasized in the first footnote of Sec.~I, the effective parameters 
$a_i$ in general cannot be defined for the $B\to VV$ decay as its amplitude 
involves more than one Lorentz scalar. In the absence of information
for the nonfactorizable contributions to various Lorentz scalars, we shall
assume generalized factorization. Under this hypothesis, 
the decay amplitude of $B\to\phi K^*$ has a similar expression as
that of $B\to\phi K$. Its decay rate is given by
\be
\Gamma(B^-\to\phi K^{*-}) &=& {p_c\over 8\pi m_B^2}\,\Bigg| {G_F\over\sqrt{2}}
V_{tb}V_{ts}^*\,f_\phi m_\phi(m_B+m_{K^*})A_1^{BK^*}(m_\phi^2)   \non\\
&\times& \left( a_3+a_4+a_5-{1\over 2}(a_7+a_9+a_{10})\right)\Bigg|^2
\left[(a-bx)^2+2(1+c^2y^2)\right],
\en
with
\be
&& x=\,A_2^{BK^*}(m_\phi^2)/A_1^{BK^*}(m_\phi^2),  \qquad \quad
y=\,V^{BK^*}(m_\phi^2)/A_1^{BK^*}(m_\phi^2),   \non \\
&& a=\,{m_B^2-m_{K^*}^2-m^2_\phi\over 2m_{K^*}m_\phi}, \qquad
b={2p_c^2m_B^2\over m_{K^*}m_\phi(m_B+m_{K^*})^2},   \qquad
c={2p_cm_B\over (m_B+m_{K^*})^2},
\en
where we have neglected contributions proportional to $X^{(B,K^*\phi)}$.

   We calculate the decay rates using two different sets of values for 
form factors:
\be
A_1^{BK^*}(0)=0.328,  \qquad A_2^{BK^*}(0)=0.331,  \qquad V^{BK^*}(0)=
0.369
\en
from \cite{BSW87} and
\be
A_1^{BK^*}(0)=0.26,  \qquad A_2^{BK^*}(0)=0.23,  \qquad V^{BK^*}(0)=
0.32
\en
from \cite{CCH}. As for the $q^2$ dependence, light-front calculations
indicate a dipole behavior for $V(q^2),~A_2(q^2)$ and a monopole dependence
for $A_1(q^2)$ \cite{CCH}. The result is shown in Fig.~5. It is interesting
to note that the branching ratios are very insensitive to the choice of
the values for form factors, (4.8) or (4.9).
We see that the allowed region is $0.7\gsim 1/\nc(V+A)
\gsim 0.25$ or $4\gsim \nc(V+A)\gsim 1.4$,
bearing in mind that this constraint is subject to the corrections from 
annihilation terms. This seems to be in contradiction to the constraint
$\nc(V+A)>4.3$ derived from $B^\pm\to\phi K^\pm$. In fact, it is expected 
in the factorization approach that
$\Gamma(B\to\phi K^*)\approx\Gamma(B\to\phi K)$ 
when the $W$-annihilation type of contributions is neglected. The current
CLEO measurements (4.3) and (4.5) are obviously not consistent with the
prediction based on factorization. One possibility is that
generalized factorization is not applicable to $B\to VV$.
Therefore, the discrepancy between ${\cal B}(B\to\phi K)$ and 
${\cal B}(B\to\phi K^*)$ will measure
the degree of deviation from the generalized factorization that has been
applied to $B\to\phi K^*$. At any rate, 
in order to clarify this issue and to pin down the effective number of 
colors $\nc(V+A)$, we need 
measurements of $B\to\phi K$ and $B\to\phi K^*$, especially the neutral 
modes, with sufficient accuracy. 

   Since CLEO has measured $B^-\to\pi^-\overline{K}^0$ and $\overline{B}^0\to
\pi^+ K^-$\cite{CLEOpik}, we have also studied these two decay modes. We 
found that for a fixed $\nc(V-A)=2$, the predicted 
branching ratios of $B\to\pi K$ 
are in agreement with the CLEO measurement within errors for all values of
$1/\nc(V+A)$. Hence, no useful constraint on $\nc(V+A)$ can be derived
from $B^-\to\pi^-\overline{K}^0$ and $\overline{B}^0\to\pi^+ K^-$.

Since $\nc(V+A)>\nc(V-A)$, one may wonder if the leading $1/N_c$ expansion may
happen to be applicable again to the matrix elements of $(V-A)(V+A)$ 
operators.
We believe that $\nc(V+A)=\infty$ is very unlikely for two reasons. First,
it will predict a too small branching ratio of $B\to\phi K^*$ as shown
in Fig.~5. Second, it implies a nonfactorizable term 
$\chi_B(V+A)\sim -{1\over 3}$, as in the charm 
case. Since the energy release in the energetic two-body decays of the 
$B$ meson is much larger than that in charm decay, it is thus expected that
\be
|\chi(D\to\overline{K}\pi)|>|\chi_B(V+A)|\sim |\chi(B\to D\pi)|.
\en
Because $\nc(V+A)\gsim 4.3$, it is then plausible to assume that 
$\chi_B(V+A)\sim -\chi(B\to D\pi)\approx -(0.10-0.12)$. Hence, $\nc(V+A)\sim
(4.3-4.9)$.

\section{Implications on charmless $B$ decays into $\eta'$ and $\eta$}
     When the preliminary CLEO measurement of $B^\pm\to\eta' K^\pm$ was 
reported  last year \cite{Smith97}
\be
{\cal B}(B^\pm\to\eta' K^\pm)=\left( 7.8^{+2.7}_{-2.2}\pm 1.0\right)
\times 10^{-5},
\en
it has stimulated a great interest in the community since 
early theoretical estimates of the $B^\pm\to\eta' K^\pm$ branching ratio 
\cite{Chau1,Du,Kramer} lie in the range of $(1-2)\times 
10^{-5}$.\footnote{The prediction ${\cal B}(B^\pm\to\eta' K^\pm)=3.6\times 
10^{-5}$ given in \cite{Chau1}
is too large by about a factor of 2 because the normalization constant 
of the $\eta'$ wave function was not taken into account in the 
form factor $F_0^{B\eta'}$. This negligence was also erroneously made in 
some recent papers on $B\to\eta' K$. Note that all early calculations
\cite{Chau1,Du,Kramer} did not take into account the anomaly contribution
to the matrix element $\la \eta'|\bar s\gamma_5s|0\ra$ (see below).}
Since then, many theoretical studies and speculation have surged, as 
evidenced by the recent literature 
\cite{Ali,Deshpande1,Halperin,CT97a,Datta,Kagan,Dighe,Fritsch,Ahmady,Du2,Ali2,He,Petrov} 
that offer various interpretations on the abnormally 
large branching ratios. It was soon realized \cite{Ali,Deshpande1} that 
the running strange quark mass at the scale
$\mu={\cal O}(m_b)$ and SU(3) breaking in the decay constants of the $\eta_0$
and $\eta_8$ will provide a large enhancement to the decay rate of $B\to\eta' 
K$ (for a review, see \cite{Chengp4}). Unfortunately, as pointed
out in \cite{Ali}, this enhancement is partially washed out by the anomaly
contribution to the matrix element $\la \eta'|\bar s\gamma_5 s|0\ra$,
an effect overlooked previously. As a consequence, the branching
ratio of $B\to\eta' K$ is of order $(2-3)\times 10^{-5}$ in the range
$0\leq 1/\nc\leq 0.5$. The discrepancy between theory and current
measurements \cite{CLEOeta'}
\be
{\cal B}(B^\pm\to\eta' K^\pm) &=& \left(6.5^{+1.5}_{-1.4}\pm 0.9\right)\times
10^{-5}, \non \\
 {\cal B}(B^0\to\eta' K^0) &=& \left(4.7^{+2.7}_{-2.0}\pm 0.9
\right)\times 10^{-5},
\en
seems to call for some new mechanisms unique to the $\eta'$ production or 
even some new physics beyond the Standard Model.

  All the previous analyses of $B\to \eta' K$ in the literature are based
on the assumption that $(\nc)_i$ are the same for $i=1,2,\cdots,10$. In this
Section we will show that the fact that $\nc(V+A)$ and $\nc(V-A)$
are not the same and that they are subject to the constraints
(3.17) and (4.10) will lead to a significant enhancement for the decay
rate of $B\to \eta' K$ at small values of $1/\nc$. Moreover, we shall
see that the prediction of ${\cal B}(B\to\eta' K)$ 
is compatible with experiment. Especially, 
the measurement of $B^0\to\eta' K^0$ is well 
explained, implying that no new mechanism in the
Standard Model or new physics beyond the Standard Model
is needed to account for the data.

   To begin with, we write down the factorizable amplitude
\be
A(B^-\to\eta' K^-) &=& {G_F\over\sqrt{2}}\Bigg\{ V_{ub}V_{us}^*\left(a_1
X^{(B\eta',K)}+a_2X^{(BK,\eta')}_u+a_1X^{(B,\eta'K)}
\right)+V_{cb}V_{cs}^*a_2X^{(BK,\eta')}_c \non \\
&-& V_{tb}V_{ts}^*\Bigg[ \left(a_4+a_{10}+2(a_6+a_8){m_K^2\over (m_s+m_u)(m_b
-m_u)}\right)X^{(B\eta',K)}   \non \\
&+& \left(2a_3-2a_5-{1\over 2}a_7+{1\over 2}a_9\right)X^{(BK,\eta')}_u 
+(a_3-a_5-a_7+a_9)X_c^{(BK,\eta')}  \non \\
&+& \left(a_4+a_{10}+2(a_6+a_8){m_B^2\over (m_s-m_u)(m_b+m_u)}\right)X^{
(B,\eta'K)}   \non \\
&+& \Bigg(a_3+a_4-a_5+{1\over 2}a_7-{1\over 2}a_9-{1\over 2}a_{10} \non\\
&+& (a_6-{1\over 2}a_8){m^2_{\eta'}\over m_s(m_b-m_s)}\left(1-{f_{\eta'}^u
\over f_{\eta'}^s}\right)\Bigg)X^{(BK,\eta')}_s   
\Bigg]\Bigg\},
\en
for $B^-\to\eta' K^-$, and
\be
A(\overline{B}^0\to\eta' \overline{K}^0) &=& {G_F\over\sqrt{2}}\Bigg\{ 
V_{ub}V_{us}^*a_2X^{(BK,\eta')}_u
+V_{cb}V_{cs}^*a_2 X^{(BK,\eta')}_c \non \\
&-& V_{tb}V_{ts}^*\Bigg[ \left(a_4-{1\over 2}a_{10}+(2a_6-a_8){m_K^2\over 
(m_s+m_d)(m_b-m_d)}\right)X^{(B\eta',K)}   \non \\
&+& \left(2a_3-2a_5-{1\over 2}a_7+{1\over 2}a_9\right)X^{(BK,\eta')}_u 
+(a_3-a_5-a_7+a_9)X_c^{(BK,\eta')}   \non \\
&+& \left(a_4+a_{10}+2(a_6+a_8){m_B^2\over (m_s-m_d)(m_b+m_d)}\right)X^{
(B,\eta'K)}   \non \\
&+& \Bigg(a_3+a_4-a_5+{1\over 2}a_7-{1\over 2}a_9-{1\over 2}a_{10} \non\\
&+& (a_6-{1\over 2}a_8){m^2_{\eta'}\over m_s(m_b-m_s)}\left(1-{f_{\eta'}^u
\over f_{\eta'}^s}\right)\Bigg)X^{(BK,\eta')}_s  
\Bigg]\Bigg\},
\en
for $\overline{B}^0\to\eta' \overline{K}^0$, where
\be
X^{(B\eta',K)} &\equiv& \la K^-|(\bar su)_\vma|0\ra\la\eta'|(\bar ub)_
\vma|B^-\ra  =\la \overline{K}^0|(\bar sd)_\vma|0\ra\la\eta'|(\bar 
db)_\vma|\bar B^0\ra \non\\
&=& if_K(m_B^2-m^2_{\eta'})F_0^{B\eta'}(m_K^2),   \non \\
X^{(BK,\eta')}_q &\equiv& \la \eta'|(\bar qq)_\vma|0\ra\la K^-|(\bar sb)_
\vma|B^-\ra= \la \eta'|(\bar qq)_\vma|0\ra \la\overline{K}^0|(\bar sb)_
\vma|\bar B^0\ra   \non \\
&=& if_{\eta'}^q(m_B^2-m^2_K)F_0^{BK}(m_{\eta'}^2),   \non \\
X^{(B,\eta' K)} &\equiv& \la\eta'K^-|(\bar su)_\vma|0\ra\la 0|(\bar ub)_
\vma|B^-\ra,
\en
and use of the isospin relation $X_d^{(BK,\eta')}=X_u^{(BK,\eta')}$ has
been made. For the amplitude of $B^-\to\eta' K^-$, the terms proportional
to $X^{(B,\eta' K)}$ and $X_c^{(BK,\eta')}$ with penguin coefficients are
often missed or not considered in previous analyses. Note that the neutral 
mode $\overline{B}^0\to\eta'\overline{K}^0$ differs from the charged
mode that it does not receive contributions from external $W$-emission and
$W$-annihilation diagrams.
From the relevant quark mixing angles 
\be
&& V_{ub}V_{us}^*=A\lambda^4(\rho-i\eta), \quad V_{cb}V_{cs}^*=A\lambda^2(1-
{1\over 2}\lambda^2), \non \\
&& V_{tb}V_{ts}^*=-A\lambda^2+{1\over 2}A(1-2\rho)\lambda^4+i\eta A\lambda^4,
\en
it is clear that $B\to\eta' K$ decays are dominated by penguin diagrams.

   The presence of the term $1-(f_{\eta'}^u/f_{\eta'}^s)$ in (5.3) and
(5.4) is necessary and mandatory in order to ensure a correct chiral-limit
behavior for the $(S-P)(S+P)$ matrix elements of the penguin operators
$O_{5,6,7,8}$. In the chiral limit $m_u,~m_d,~m_s\to 0$, the ratio
$m_K^2/(m_s+m_u)=m^2_\pi/(m_u+m_d)$ remains finite\footnote{For the 
annihilation term, the chiral-limit behavior of ${m_B^2\over m_b(m_s-m_u)}
X^{(B,\eta'K)}$ is supposed to be taken care of by the form factors in 
$X^{(B,\eta' K)}$.},
but this is no longer the
case for $m_{\eta'}^2/m_s$ associated with the matrix element
$\la\eta'|\bar s\gamma_5s|0\ra$ since the $\eta'$ mass originates from
the QCD anomaly and does not vanish
in the chiral limit. As pointed out in \cite{Kagan,Ali}, due to
the presence of the anomaly in the equation of motion
\be
\partial^\mu(\bar s\gamma_\mu\gamma_5 s)=2m_s\bar si\gamma_5 s+{\alpha_s
\over 4\pi}G_{\mu\nu}\tilde{G}^{\mu\nu},
\en
it is erroneous to apply the relation
\be
\la\eta'|\bar s\gamma_5 s|0\ra=-i{m_{\eta'}^2\over 2m_s}\,f_{\eta'}^s,
\en
as adopted previously in the literature,
where $\la 0|\bar q\gamma_\mu\gamma_5 q|\eta'\ra=if_{\eta'}^q p_\mu$. 
Neglecting the $u$ and $d$ quark masses in the equations of motion
leads to \cite{Ball}
\be
\la \eta' |{\alpha_s\over 4\pi}G\tilde G|0\ra=f_{\eta'}^u m^2_{\eta'}
\en
and hence \cite{Kagan,Ali}
\be
\la\eta'|\bar s\gamma_5 s|0\ra=-i{m_{\eta'}^2\over 2m_s}\,\left(f_{\eta'}^s
-f^u_{\eta'}\right).
\en
It is easily seen that this matrix element 
has the correct chiral behavior. It should be stressed that in order to
go the chiral-symmetry limit, one must consider both $m_s\to 0$ and $\theta
\to 0$ together \cite{Akhoury}, where $\theta$ is the $\eta-\eta'$ mixing 
angle to be defined below.
Since $f_{\eta'}^u\sim {1\over 2}f_{\eta'}^s$ (see below) and the decay
amplitude is dominated by $(S-P)(S+P)$ matrix elements, it is obvious that
the decay rate of $B\to\eta' K$ is reduced considerably by the
presence of the anomaly term in $\la\eta'|\bar s\gamma_5 s|0\ra$.

   To determine the decay constant $f_{\eta'}^q$, we need to know the
wave functions of the physical $\eta'$ and $\eta$ states which are related to
that of the SU(3) singlet state $\eta_0$ and octet state $\eta_8$ by
\be
\eta'=\eta_8\sin\theta+\eta_0\cos\theta, \qquad \eta=\eta_8\cos\theta-\eta_0
\sin\theta,
\en
with $\theta\approx -20^\circ$. When the $\eta-\eta'$ mixing angle is 
$-19.5^\circ$, 
the $\eta'$ and $\eta$ wave functions have simple expressions \cite{Chau1}:
\be
|\eta'\ra={1\over\sqrt{6}}|\bar uu+\bar dd+2\bar ss\ra, \qquad
|\eta\ra={1\over\sqrt{3}}|\bar uu+\bar dd-\bar ss\ra,
\en
recalling that
\be
|\eta_0\ra={1\over\sqrt{3}}|\bar uu+\bar dd+\bar ss\ra, \qquad
|\eta_8\ra={1\over\sqrt{6}}|\bar uu+\bar dd-2\bar ss\ra.
\en
At this specific mixing angle, $f_{\eta'}^u={1\over 2}f_{\eta'}^s$ in the
SU(3) limit. Introducing the decay constants $f_8$ and $f_0$ by 
\be
\la 0|A_\mu^0|\eta_0\ra=if_0 p_\mu, \qquad \la 0|A_\mu^8|\eta_8\ra=if_8 p_\mu,
\en
then $f_{\eta'}^u$ and $f_{\eta'}^s$ are related to $f_8$ and $f_0$ by
\footnote{A two-mixing-angle parametrization of the $\eta$ and $\eta'$
wave functions: $\eta'=\eta_8\sin\theta_8+\eta_0\cos\theta_0,~\eta=\eta_8
\cos\theta_8-\eta_0\sin\theta_0$,
is employed in \cite{Ali} for the calculation of $B\to\eta'(\eta) K$.
However, in the absence of mixing with other pseudoscalar mesons,
this parametrization will destroy the orthogonality of the 
physical states $\eta$ and $\eta'$ if $\theta_0\neq \theta_8$. 
Due to SU(3) breaking the matrix elements $\la 0|A_\mu^{0(8)}|\eta_{8(0)}
\ra$ do not vanish in general and they will induce a two-angle mixing 
among the decay constants:
\be
f_{\eta'}^u={f_8\over\sqrt{6}}\sin\theta_8+{f_0\over\sqrt{3}}\cos\theta_0,
\qquad f_{\eta'}^s=-2{f_8\over\sqrt{6}}\sin\theta_8+{f_0\over\sqrt{3}}\cos
\theta_0.     \non
\en
Based on the ansatz that the decay constants in the quark flavor basis
follow the pattern of particle state mixing,
relations between $\theta_8,~\theta_0$ and $\theta$
are derived in \cite{Kroll2}, where $\theta$ is the $\eta-\eta'$ mixing angle
introduced in (5.11). It is found in \cite{Kroll2} that
phenomenologically $\theta_8=-21.2^\circ,~\theta_0=-9.2^\circ$ and
$\theta=-15.4^\circ$. It must be accentuated that the  
two-mixing angle formalism proposed in \cite{Leutwyler,Kroll2} 
applies to the decay constants of the $\eta'$ and $\eta$ rather than to
their wave functions. Numerically, we find that the branching ratios shown
in Table I (see below) calculated in one-angle and two-angle 
mixing schemes are 
different by at most 7\%. In the present paper we shall
employ the former scheme.}
\be
f_{\eta'}^u={f_8\over\sqrt{6}}\sin\theta+{f_0\over\sqrt{3}}\cos\theta,
\qquad f_{\eta'}^s=-2{f_8\over\sqrt{6}}\sin\theta+{f_0\over\sqrt{3}}\cos
\theta.
\en
Likewise, for the $\eta$ meson
\be
f_{\eta}^u={f_8\over\sqrt{6}}\cos\theta-{f_0\over\sqrt{3}}\sin\theta,
\qquad f_{\eta}^s=-2{f_8\over\sqrt{6}}\cos\theta-{f_0\over\sqrt{3}}\sin\theta.
\en

   The factorizable amplitude denoted by $X_c^{(BK,\eta')}$ involves a 
conversion of the $c\bar c$ pair into the $\eta'$ via two gluon exchanges.
Although the charm content of the $\eta'$ is {\it a priori} expected to be 
small, its contribution is potentially important because the CKM mixing
angle $V_{cb}V_{cs}^*$ is of the same order of magnitude as that of the
penguin amplitude [cf. Eq.~(5.6)] and yet its effective coefficient $a_2$ 
is larger than the penguin coefficients by an order of magnitude. The 
decay constant $f_{\eta'}^c$, 
defined by $\la 0|\bar c\gamma_\mu\gamma_5c|\eta'\ra=if_{\eta'}^c
q_\mu$, has been estimated to be $f_{\eta'}^c=(50-180)$ MeV,
based on the OPE, large-$N_c$ approach and QCD low energy 
theorems \cite{Halperin}. It was claimed in \cite{Halperin} 
that $|f_{\eta'}^c|\sim 140$ MeV is needed in order to exhaust the 
CLEO observation of $B^\pm\to \eta' K^\pm$ and $B\to\eta'+X$ by the mechanism
$b\to c\bar c+s\to\eta'+s$ via gluon exchanges. However, a large value of 
$f_{\eta'}^c$ seems to be ruled out for several reasons \cite{Chengp4}.
For example, from the
data of $J/\psi\to\eta_c\gamma$ and $J/\psi\to\eta' \gamma$, one can show
that $|f_{\eta'}^c|\geq 6$ MeV, where the lower bound corresponds to the 
nonrelativistic quark model estimate. Based 
on the $\eta\gamma$ and $\eta'\gamma$ transition form factor data, the range 
of allowed $f_{\eta'}^c$ was recently estimated to be $-65\,{\rm MeV}
\leq f_{\eta'}^c\leq 15$ MeV \cite{Kroll}. A most recent reevaluation
of $f_{\eta'}^c$ along the line of \cite{Halperin} yields $f_{\eta'}^c
=-(12.3\sim 18.4)$\,MeV \cite{Araki}, which is in strong contradiction 
in magnitude and sign to the estimate of \cite{Halperin}.
The sign of $f_{\eta'}^c$ can be fixed by using QCD anomaly and
is found to be negative \cite{Ali2} (see also \cite{Petrov,Kroll2,Araki}). 
In the presence of the charm content in the $\eta_0$, an additional mixing
angle $\theta_c$ is needed to be introduced:
\be
|\eta_0\ra &=& {1\over \sqrt{3}}\cos\theta_c|u\bar u+d\bar d+s\bar s\ra
+\sin\theta_c|c\bar c\ra,   \non \\
|\eta_c\ra &=& -{1\over \sqrt{3}}\sin\theta_c|u\bar u+d\bar d+s\bar s\ra
+\cos\theta_c|c\bar c\ra.
\en
Then $f_{\eta'}^c=\cos\theta\tan\theta_c f_{\eta_c}$ and 
$f_\eta^c=-\sin\theta\tan\theta_cf_{\eta_c}$,
where the decay constant $f_{\eta_c}$ can be extracted from $\eta_c\to
\gamma\gamma$, and $\theta_c$ from $J/\psi\to\eta_c\gamma$ and 
$J/\psi\to\eta'\gamma$ \cite{Ali}. In the present paper we shall use
\be
f_{\eta'}^c=-6\,{\rm MeV},  \qquad f_\eta^c=-\tan\theta f_{\eta'}^c=
-2.4\,{\rm MeV},
\en
for $\theta=-22^\circ$ (see below), which are very close to the values
\be
f_{\eta'}^c=-(6.3\pm 0.6)\,{\rm MeV},  \qquad f_\eta^c=-(2.4\pm 0.2)\,{\rm 
MeV}
\en
obtained in \cite{Kroll2}.

   Using $F_0^{BK}(0)=0.34$ \cite{CCH}, $\sqrt{3}F_0^{B\eta_0}(0)=0.33$
\footnote{The form factors $F_0^{B\eta'}(0)=0.254$ and $F_0^{B\eta}(0)=
0.307$ given in \cite{BSW87} do not take into account the wave function
normalization of the physical $\eta'$ and $\eta$ states. Since it is not clear
to us what is the $\eta-\eta'$ mixing angle employed in \cite{BSW87}, we shall
follow \cite{Ali,Deshpande1} to use the nonet symmetry relation 
$\sqrt{3}F_0^{B\eta_0}(0)
=\sqrt{6}F_0^{B\eta_8}(0)=F_0^{B\pi^\pm}(0)\approx 0.33$ to obtain 
$F_0^{B\eta_0}$, $F_0^{B\eta_8}$ and hence the form factors $F_0^{B\eta'}$
as well as $F_0^{B\eta}$ for a given $\theta$.}
for form factors, $f_0=f_8=f_\pi$, $m_s(1\,{\rm GeV})=150$ MeV,
$\theta=-19.5^\circ$ and Eq.~(5.10) for the matrix element 
$\la\eta'|\bar s\gamma_5 s|0\ra$, we find that ${\cal B}(B\to\eta' K)=
(0.9- 1.0)\times 10^{-5}$ and it is insensitive to $\nc$ and the choice of 
Wolfenstein parameters $\rho$ and $\eta$ so long as $\sqrt{\rho^2+\eta^2}
\approx 0.37$,  where
$\nc=\nc(V+A)=\nc(V-A)$. The discrepancy between theory and experiment can be
greatly improved by the accumulation of several enhancements. First of all, 
the running quark masses appearing in the $(S-P)(S+P)$ matrix elements
should be applied at the scale $\mu={\cal O}(m_b)$ as given in 
Eq.~(2.13) so that the $(S-P)(S+P)$ matrix element is enhanced due to
the decrease of $m_s(\mu)$ at $\mu=m_b$. (The sensitivity of the 
branching ratio to $m_s$ was first noticed in \cite{Kagan}.)
Second, a recent analysis of the data of $\eta,\eta'\to \gamma\gamma$
and $\eta,\eta'\to\pi\pi\gamma$ yields \cite{Holstein}
\be
{f_8\over f_\pi}=1.38\pm 0.22, \qquad {f_0\over f_\pi}=1.06\pm 0.03,
\qquad \theta=-22.0^\circ\pm 3.3^\circ,
\en
implying some SU(3) breaking in the decay constants. Applying the new 
values of the aforementioned parameters,
the result for the branching ratio of $B^\pm\to\eta' K^\pm$ 
is shown in Fig.~6 vs $1/\nc$ (see the lower set of solid and dotted
curves). We find that ${\cal B}(B^\pm\to\eta' K^\pm)$ is enhanced from
$(0.9-1.0)\times 10^{-5}$ to $(2-3)\times 10^{-5}$. The latter result is
in agreement with \cite{Ali} (see the lower set of curves with 
negative $f_{\eta'}^c$ in Fig.~17 of \cite{Ali}).
The enhancement is due mainly to the running strange quark
mass at $\mu=m_b$ and SU(3) breaking effects in the decay constants
$f_0$ and $f_8$. From Fig.~6 we see that (i) in the absence of the
anomaly contribution to $\la\eta'|\bar s\gamma_5 s|0\ra$, the 
branching ratios (the upper set of solid and dotted curves) will be
further enhanced in a sizable way
(of course, it is erroneous to neglect such an anomaly effect), and (ii)
the contribution of $c\bar c$ conversion into the $\eta'$ becomes
destructive when $1/\nc<0.28$. This is understandable because $a_2$
becomes negative at small values of $1/\nc$ so that the term 
$a_2X_c^{(BK,\eta')}$ contributes in opposite sign to the penguin amplitudes.
Therefore, the charm content of the $\eta'$ is not welcome for explaining
${\cal B}(B\to\eta' K)$ at small $1/\nc$.

   Thus far it has been assumed in the analysis of $B\to\eta' K$ that the
nonfactorizable effects lumped into $a_i$ via $(\nc)_i$ are the same for
$i=1,2,\cdots,10$. However, we have pointed out in Sec.~III that
$\nc(V-A)$ in hadronic charmless $B$ decays is most likely very 
similar to that in $B\to D\pi$, namely $\nc(V-A)\sim\nc(B\to D\pi)\approx
2$. In fact, we just showed that the charm content of the $\eta'$
will make the discrepancy between theory and experiment even worse 
at small values of $1/\nc$ if $\nc(V-A)$ is the same
as $\nc(V+A)$. Setting $\nc(V-A)=2$, we find that (see Figs.~7 and 8)
the decay rates of $B\to\eta' K$ are considerably enhanced especially
at small $1/\nc(V+A)$. That is, ${\cal B}(B^\pm\to\eta' K^\pm)$ at 
$1/\nc(V+A)\leq 0.2$ is enhanced from $(2.5-3)\times 10^{-5}$ to $(3.7-5)
\times 10^{-5}$. 
First, the $\eta'$ charm content contribution
$a_2X_c^{(BK,\eta')}$ now always contributes in the right direction to
the decay rate irrespective of the value of $\nc(V+A)$. Second, the 
interference in
the spectator amplitudes of $B^\pm\to\eta' K^\pm$ is constructive. 
Third, the term proportional to
\be
2(a_3-a_5)X_u^{(BK,\eta')}+(a_3+a_4-a_5)X_s^{(BK,\eta')}
\en
in Eqs.~(5.3) and (5.4) is enhanced when $(\nc)_3=(\nc)_4=2$. It is evident
from Fig.~8 that the measurement of $\overline{B}^0\to\eta'\overline{K}^0$
is well explained in the present framework based on the Standard Model 
within the allowed range
$1/\nc(V+A)\lsim 0.23$ extracted from $B^\pm\to\phi K^\pm$. Contrary to some
early claims, we see that it is not necessary to invoke some new mechanisms,
say the SU(3)-singlet contribution $S'$ \cite{Dighe}, to explain the
data. The agreement with
experiment provides another strong support for $\nc(V-A)\sim 2$ and for
the relation $\nc(V+A)>\nc(V-A)$.
As for the decay $B^\pm\to\eta' K^\pm$, the predicted branching ratio,
say $4\times 10^{-5}$ at our preferred value $\nc(V+A)\sim 5$ 
(see Table I), is compatible with the data, though it is on the lower
side. For a slightly enhanced decay constant $f_{\eta'}^c\approx -15$ MeV, 
as implied
by a recent theoretical estimate \cite{Araki}, we obtain ${\cal B}(B\to
\eta' K)=(4.6-5.9)\times 10^{-5}$ at $1/\nc(V+A)\leq 0.2$, which agrees
with experiment very nicely.
Note that the CLEO data of $B^\pm\to\eta' K^\pm$ and $B^0\to\eta' K$
are in good agreement within one sigma error [see (5.2)], though the 
charged mode is more reliable. It is conceivable that when errors are
improved and refined, the two values will converge eventually.

   We have also studied the decays $B\to\eta K,~\eta' K^*,~\eta K^*$.
The decay amplitude of $B\to\eta K$ is the same as $B\to\eta' K$ except
for a trivial replacement of the index $\eta'$ by $\eta$.
As a general rule, the factorizable amplitude of $B\to\eta^{(')}K^*$
can be obtained from the $B\to\eta^{(')}K$ one by (i) replacing
the term $m^2_P/[(m_1+m_2)(m_3-m_4)]$ by $-m^2_P/[(m_1+m_2)(m_3+m_4)]$
and the index $K$ by $K^*$,
and (ii) discarding the $(S-P)(S+P)$ contribution associated with 
$X^{(B\eta^{(')},
K^*)}$. For example, the decay amplitude of $B^-\to\eta' K^{*-}$ can be easily
read from (5.3) to be:
\be
A(B^-\to\eta' K^{*-}) &=& {G_F\over\sqrt{2}}\Bigg\{ V_{ub}V_{us}^*\left(a_1
X^{(B\eta',K^*)}+a_2X^{(BK^*,\eta')}_u+a_1X^{(B,\eta'K^*)}
\right)+V_{cb}V_{cs}^*a_2X^{(BK^*,\eta')}_c \non \\
&-& V_{tb}V_{ts}^*\Bigg[ (a_4+a_{10})X^{(B\eta',K^*)}   
+ \Bigg(a_3+a_4-a_5+{1\over 2}a_7-{1\over 2}a_9-{1\over 2}a_{10} \non \\
&-& (a_6-{1\over 2}a_8){m^2_{\eta'}\over m_s(m_b+m_s)}\left(1-{f_{\eta'}^u
\over f_{\eta'}^s}\right)\Bigg)X^{(BK^*,\eta')}_s   \non \\
&+& \left(2a_3-2a_5-{1\over 2}a_7+{1\over 2}a_9\right)X^{(BK^*,\eta')}_u 
+(a_3-a_5-a_7+a_9)X_c^{(BK^*,\eta')}  \non \\
&+& \left(a_4+a_{10}-2(a_6+a_8){m_B^2\over (m_s+m_u)(m_b+m_u)}\right)X^{
(B,\eta'K^*)}\Bigg]\Bigg\},
\en
with
\be
X^{(B\eta',K^*)} &\equiv& \la K^{*-}|(\bar su)_\vma|0\ra\la\eta'|(\bar ub)_
\vma|B^-\ra  \non \\
&=& -2f_{K^*}m_{K^*}F_1^{B\eta'}(m_{K^*}^2)(\vp\cdot p_B),   \non \\
X^{(BK^*,\eta')}_q &\equiv& \la \eta'|(\bar qq)_\vma|0\ra\la K^{*-}|(\bar sb)_
\vma|B^-\ra   \non \\
&=& -2f_{\eta'}^qm_{K^*}A_0^{BK^*}(m_{\eta'}^2)(\vp\cdot p_B),   \non \\
X^{(B,\eta' K^*)} &\equiv& \la\eta'K^{*-}|(\bar su)_\vma|0\ra\la 0|(\bar ub)_
\vma|B^-\ra.
\en
From Table I we see that the electroweak penguin is generally 
small due to the smallness of its  Wilson coefficients, but it does play an
essential role in the decays $B^\pm\to\eta K^\pm$ and $B^0\to\eta K^0$. 
It is interesting to note that the branching ratios of 
$B\to\eta^{(')} K^{(*)}$ are all less than $1\times 10^{-5}$ except
for $B\to\eta' K$, which has a very large branching ratio, of order $(4-6)
\times 10^{-5}$. It has been argued in \cite{Halperin} that 
${\cal B}(B\to\eta' K^*)$ is about twice larger than that
of $B\to\eta' K$, which is certainly not the case in our calculation. 
The ratios of various decay rates are predicted to be
\be
{ {\cal B}(B\to\eta' K)\over {\cal B}(B\to\eta K)}=\cases{ 72 \cr 296 \cr},
\qquad { {\cal B}(B\to\eta' K^*)\over {\cal B}(B\to\eta K^*)}=\cases{ 0.06 
& charged~$B$;  \cr 0.02 &neutral~$B$,  \cr}
\en
for positive $\rho$.
The destructive (constructive) interference between the terms 
$X^{(B\eta^{(')},K)}$
and $a_6X_s^{(BK,\eta^{(')})}$ explains the ratio ${\cal B}(B\to\eta' K)/ 
{\cal B}(B\to\eta K)$: $X_s^{(BK,\eta')}$ has a sign opposite to
$X_s^{(BK,\eta)}$ as one can easily see from the wave functions of
the $\eta$ and $\eta'$, Eq.~(5.11). Since the sign of 
$a_6X_s^{(BK^*,\eta^{(')})}$ is flipped in $B\to\eta^{(')} K^*$ decays, the
interference effect becomes the other way around: constructive in $B\to
\eta K^*$ and destructive in $B\to\eta' K^*$.

\vskip 0.4cm
{\footnotesize
\begin{table}[ht]
Table I. Branching ratios averaged over CP-conjugate modes
for charmless $B$ decays to the
$\eta'$ and $\eta$, where ``Tree" refers to branching ratios from 
tree diagrams only, ``Tree+QCD" from tree and QCD penguin diagrams, and 
``Full" denotes full contributions from tree, QCD and electroweak (EW) penguin
diagrams in conjunction with contributions from the process $c\bar 
c\to\eta_0$. Predictions are for $k^2=m_b^2/2$, 
$\eta=0.35,~\rho=-0.12$ (the first number in parentheses) and $\eta=0.34,~\rho
=0.16$ (the second number in parentheses). The decay constants $f_{\eta'}^c
=-6$ MeV and $f_\eta^c=-2.4$ MeV are used. The effective number of colors
is taken to be $\nc(V-A)=2$ and $\nc(V+A)=5$.
The running quark masses at $\mu=m_b$ are given by (2.13).
\begin{center}
\begin{tabular}{|l|c c c c |c|} 
Decay & Tree & Tree$+$QCD & Tree$+$QCD$+$EW & Full & Expt. \cite{CLEOeta'} \\ 
\hline
$B^\pm\to\eta' K^\pm$ & $1.48\times 10^{-7}$ & $(3.56,~3.33)\,10^{-5}$ & 
$(3.42,~3.20)\,10^{-5}$ & $(3.99,~3.74)\,10^{-5}$ &
$(6.5^{+1.5}_{-1.4}\pm 0.9)\,10^{-5}$ \\
$B^\pm\to\eta K^\pm$ & $4.18\times 10^{-7}$ & $(0.59,~1.27)\,10^{-6}$ &
$(3.91,~7.10)\,10^{-7}$ & $(3.88,\,5.17)\,10^{-7}$  & $<1.4\times 10^{-5}$ \\
$B^\pm\to\eta' K^{*\pm}$ & $2.44\times 10^{-7}$ & $(3.66,~4.00)\,10^{-7}$ &
$(3.54,~4.62)\,10^{-7}$ & $(5.73,\,3.53)\,10^{-7}$ & $<13\times 10^{-5}$ \\
$B^\pm\to\eta K^{*\pm}$ & $5.98\times 10^{-7}$ & $(6.42,~4.09)\,10^{-6}$ &
$(8.30,~5.58)\,10^{-6}$ & $(9.22,\,6.32)\,10^{-6}$ & $<3.0\times 10^{-5}$ \\
$B^\pm\to\eta'\pi^{\pm}$ & $2.13\times 10^{-6}$ & $(1.47,~2.53)\,10^{-6}$ &
$(1.49,~2.51)\,10^{-6}$ & $(1.52,\,2.75)\, 10^{-6}$ & $<3.1\times 10^{-5}$ \\
$B^\pm\to\eta \pi^{\pm}$ & $6.06\times 10^{-6}$ & $(4.16,~7.11)\,10^{-6}$ &
$(4.11,~7.22)\,10^{-6}$ & $(4.14,\,7.38)\, 10^{-6}$ & $<1.5\times 10^{-5}$ \\
$B^\pm\to\eta' \rho^{\pm}$ & $4.44\times 10^{-6}$ & $(3.93,~4.69)\,10^{-6}$ &
$(3.94,~4.68)\,10^{-6}$ & $(3.87,\,4.88)\, 10^{-6}$ & $<4.7\times 10^{-5}$ \\
$B^\pm\to\eta \rho^{\pm}$ & $1.08\times 10^{-5}$ & $(0.98,~1.13)\,10^{-5}$ &
$(0.95,~1.14)\,10^{-5}$ & $(0.95,\,1.15)\, 10^{-5}$ & $<3.2\times 10^{-5}$ \\
\hline
$ B_d\to\eta' K^0$ & $5.38\times 10^{-9}$ &$(3.20,\,3.23)\, 10^{-5}$
&$(3.00,\,3.03)\, 10^{-5}$ &$(3.52,\,3.55)\, 10^{-5}$ & $(4.7^{+2.7}_{-2.0}
\pm 0.9)\, 10^{-5}$ \\
$ B_d\to\eta K^0$  & $2.05\times 10^{-8}$ & $(3.99,\,5.54)\, 10^{-7}$ 
& $(1.62,\,2.57)\,10^{-7}$ & $(0.64,\,1.20)\,10^{-7}$ & $<3.3\times 10^{-5}$\\
$ B_d\to\eta' K^{*0}$ & $4.49\times 10^{-9}$ & $(1.33,\,3.29)\,
10^{-7}$ & $(1.46,\,4.56)\, 10^{-7}$ & $(2.40,\,0.87)\, 10^{-7}$ &
$<3.9\times10^{-5}$ \\ 
$ B_d\to\eta K^{*0}$ & $1.75\times 10^{-8}$ & $(5.19,\,3.70)\,
10^{-6}$  & $(6.99,\,4.69)\, 10^{-6}$  & $(7.85,\,5.40)\, 10^{-6}$  &
$<3.0\times 10^{-5}$ \\ 
$ B_d\to\eta'\pi^{0}$ & $2.14\times 10^{-10}$ & $(1.75,\,1.10)\,
10^{-7}$  & $(1.34,\,0.85)\, 10^{-7}$  & $(1.87,\,1.27)\, 10^{-7}$  & 
$<1.1\times 10^{-5}$ \\
$ B_d\to\eta \pi^{0}$ & $1.01\times 10^{-8}$ & $(3.99,\,2.97)\,
10^{-7}$ & $(3.77,\,2.83)\, 10^{-7}$ & $(4.09,\,3.11)\, 10^{-7}$ & 
$<0.8\times 10^{-5}$ \\
$ B_d\to\eta'\rho^{0}$ & $1.34\times 10^{-8}$ &$(3.43,\,1.81)\,
10^{-8}$ &$(2.85,\,1.65)\, 10^{-8}$ &$(2.15,\,1.83)\, 10^{-8}$ & 
$<2.3\times 10^{-5}$ \\
$ B_d\to\eta \rho^{0}$ & $1.99\times 10^{-8}$ &$(4.27,\,9.07)\,
10^{-8}$ &$(3.14,\,5.84)\, 10^{-8}$ &$(3.11,\,5.36)\, 10^{-8}$ &
$<1.3\times 10^{-5}$   \\
\end{tabular}
\end{center}  
\end{table} }
\vskip 0.4cm

  To discuss the decays $B\to\eta^{(')}\pi(\rho)$, we consider $B^-\to\eta'
\pi^-$ as an illustration. Its decay amplitude is
\be
A(B^-\to\eta' \pi^-) &=& {G_F\over\sqrt{2}}\Bigg\{ V_{ub}V_{ud}^*\left(a_1
X^{(B\eta',\pi)}+a_2X^{(B\pi,\eta')}_u+2a_1X^{(B,\eta'\pi)}
\right)+V_{cb}V_{cd}^*a_2X^{(B\pi,\eta')}_c \non \\
&-& V_{tb}V_{td}^*\Bigg[ \left(a_4+a_{10}+2(a_6+a_8){m_\pi^2\over 
(m_d+m_u)(m_b-m_u)}\right)X^{(B\eta',\pi)}   \non \\
&+& \left(a_3-a_5+{1\over 2}a_7-{1\over 2}a_9\right)X^{(B\pi,\eta')}_s 
+(a_3-a_5-a_7+a_9)X_c^{(B\pi,\eta')}  \non \\
&+& \left(2a_4+2a_{10}+4(a_6+a_8){m_B^2\over (m_d-m_u)(m_b+m_u)}\right)X^{
(B,\eta'\pi)}   \non \\
&+& \Bigg(2a_3+a_4-2a_5-{1\over 2}(a_7-a_9+a_{10}) \non \\
&+& (a_6-{1\over 2}a_8){m^2_{\eta'}\over m_s(m_b-m_d)}\left({f_{\eta'}^s
\over f_{\eta'}^u}-1\right)r_{\eta'}\Bigg)X^{(B\pi,\eta')}_u   
\Bigg]\Bigg\},   \label{eta'pi}
\en
where
\be
r_{\eta'}=\,{\sqrt{2f_0^2-f_8^2}\over\sqrt{2f_8^2-f_0^2}}\,{\cos\theta+
{1\over \sqrt{2}}\sin\theta\over \cos\theta-\sqrt{2}\sin\theta},
\en
and
\be
X^{(B\eta',\pi)} &\equiv& \la \pi^-|(\bar du)_\vma|0\ra\la\eta'|(\bar ub)_
\vma|B^-\ra  
= if_\pi(m_B^2-m^2_{\eta'})F_0^{B\eta'}(m_\pi^2),   \non \\
X^{(B\pi,\eta')}_q &\equiv& \la \eta'|(\bar qq)_\vma|0\ra\la \pi^-|(\bar db)_
\vma|B^-\ra= 
 if_{\eta'}^q(m_B^2-m^2_\pi)F_0^{B\pi}(m_{\eta'}^2),   \non \\
X^{(B,\eta' \pi)} &\equiv& \la\eta'\pi^-|(\bar du)_\vma|0\ra\la 0|(\bar ub)_
\vma|B^-\ra. 
\en
In deriving (\ref{eta'pi}) we have applied the matrix elements
\footnote{The matrix element $\la\eta'|\bar u\gamma_5u|0\ra$ can be 
obtained from \cite{Kroll2}
and it is slightly different from the corresponding one 
in \cite{Akhoury,Ball}:
\be
\la\eta'|\bar u\gamma_5u|0\ra=\,{f_8\cos\theta+{1\over\sqrt{2}}f_0\sin\theta
\over f_8\cos\theta-\sqrt{2}f_0\sin\theta}\,\la\eta'|\bar s\gamma_5s|0\ra=
-{1\over 2}\,{f_\eta^s\over f_\eta^u}\,\la\eta'|\bar s\gamma_5s|0\ra. \non
\en}
\be
\la\eta'|\bar u\gamma_5u|0\ra=\la\eta'|\bar d\gamma_5d|0\ra=r_{\eta'}
\,\la\eta'|\bar s\gamma_5s|0\ra,
\en
with $r_{\eta'}$ being given by (5.26).

Since $V_{ub}V^*_{ud},~V_{cb}V^*_{cd},~V_{tb}V^*_{td}$ are all comparable 
in magnitude [cf. Eq.~(3.15)] and since the Wilson coefficients of 
penguin operators are rather small, it is expected that $B\to\eta^{(')}\pi,
\eta^{(')}\rho$ are dominated by spectator diagrams\footnote{The branching 
ratios of $B\to\eta^{(')}\pi,~\eta^{(')}\rho$ are
largely overestimated in \cite{CT97a,Chengp4} as the incorrect matrix element
$\la\eta'|\bar u\gamma_5u|0\ra=-im_{\eta'}^2f_{\eta'}^u/(2m_u)$ is applied
there.}.
From Table I we see that this is indeed the case except for the decay modes
$B^0\to\eta^{(')}\pi^0$ which are penguin dominated. 
To compute the decay rate of $B\to\eta\pi(\rho)$ we have applied the 
matrix element
$\la\eta|\bar u\gamma_5u|0\ra=r_\eta\la\eta|\bar s\gamma_5s|0\ra$ with
\be
r_{\eta}=-{1\over 2}\,{\sqrt{2f_0^2-f_8^2}\over\sqrt{2f_8^2-f_0^2}}\,
{\cos\theta-\sqrt{2}\sin\theta\over \cos\theta+{1\over\sqrt{2}}
\sin\theta}.
\en

 The mechanism of $c\bar c\to\eta_0$ is less significant in $B\to\eta^{(')}
\pi(\rho)$ decays because it does not gain advantage from the quark mixing 
angle as in the case of $B\to\eta^{(')}K(K^*)$. We see from Table I 
the minor role played by the charm content of the 
$\eta'$ except for the decay $B^0\to\eta'\pi^0$.
In general, the decay rates of $B\to\eta^{(')}\pi(\rho)$ are
not sensitive to the values of $\nc(V-A)$ and $\nc(V+A)$ and do not vary
significantly from channel to channel:
\be
{\cal B}[B^\pm\to\eta^{(')}\pi(\rho)] \sim (3-10)\times 10^{-6},   \qquad
{\cal B}[\overline{B}^0\to\eta^{(')}\pi(\rho)] \sim (0.2-4)\times 10^{-7}.
\en
It is interesting to note that ${\cal B}[B\to\eta\pi(\rho)]>{\cal B}[B\to
\eta'\pi(\rho)]$.

\section{Difficulties with $B^-\to K^-\omega$}

   Up to now  we have shown that CLEO results on hadronic charmless $B$
decays can be satisfactorily explained provided that  $\nc(V-A)\approx 2$ and
$\nc(V+A)\gsim {\cal O}(4)$.
However, there is one CLEO measurement, namely
the decay $B^\pm\to\omega K^\pm$, that is beyond our explanation and hence
may impose a potentially serious difficulty. In this Section we will 
first explore
the problem and then proceed to suggest some possible solutions.

   The decay amplitude of $B^-\to\omega K^-$ is very similar to
$B^-\to\omega \pi^-$ and has the expression
\be
A(B^-\to\omega K^-) &=& {G_F\over\sqrt{2}}\Bigg\{ V_{ub}V_{us}^*\left(a_1
X^{(B\omega,K)}+a_2X_u^{(BK,\omega)}+a_1X^{(B,K\omega)}\right)  \non \\
&-& V_{tb}V_{ts}^*\Bigg[ \left(a_4+a_{10}-2(a_6+a_8){m_K^2\over (m_b+m_u)
(m_s+m_u)}\right)X^{(B\omega,K)}   \non \\
&+& {1\over 2}\left(4a_3+4a_5+a_7+a_9\right)X_u^{(BK,\omega)}  
\non \\
&+& \left(a_4+a_{10}-2(a_6+a_8){m_B^2\over (m_b+m_u)
(m_s+m_u)}\right)X^{(B,K\omega)} \Bigg]\Bigg\},
\en
where 
\be
X^{(B\omega,K)} &\equiv& \la K^-|(\bar su)_\vma|0\ra\la\omega|(\bar ub)_\vma|
B^-\ra=-2f_K m_\omega A_0^{B\omega}(m_K^2)(\vp\cdot p_B),   \non \\
X^{(BK,\omega)}_u &\equiv& \la \omega|(\bar uu)_\vma|0\ra\la K^-|(\bar sb)_
\vma|B^-\ra=-\sqrt{2}f_\omega m_\omega F_1^{BK}(m_{\omega}^2)(\vp\cdot p_B), 
  \non \\
X^{(B,\omega K)} &\equiv& \la\omega K^-|(\bar su)_\vma|0\ra\la 0|(\bar ub)_
\vma|B^-\ra.
\en
We see from Fig.~9 that the calculated branching ratio using $\nc(V-A)=2$,
$F_1^{BK}(0)=0.34$ and $A_0^{B\omega}(0)=0.28/\sqrt{2}$ \cite{BSW87} 
is too small compared to experiment \cite{CLEOomega2}:
\be
{\cal B}(B^\pm\to\omega K^\pm)=\left(1.5^{+0.7}_{-0.6}\pm 0.2\right)
\times 10^{-5}.
\en
In fact, all the region of $1/\nc(V+A)<0.9$ is excluded. Nevertheless, if 
$\nc(V-A)$ is taken to be the same as $\nc(V+A)$, then a rather small value of
$1/\nc<0.05$ is experimentally allowed \cite{Ali,Deshpande1} (see Fig.~10).
In other words, $\nc$ is preferred to be very large in $B^\pm\to\omega K^\pm$. 
In our opinion, however, a very large value of $\nc(V-A)$ is rather 
unlikely for several reasons: (i)
A small $\nc(V-A)\approx 2$ is favored in other charmless $B$ decays:
$B\to\pi\pi,~\pi\omega$ and $B\to\eta' K$. (ii) It will lead to a too
large nonfactorizable term, which is not consistent with the small 
nonfactorizable 
effect observed in the spectator amplitudes of $B\to D\pi$ and the picture
that the nonperturbative feature of nonfactorizable effects is loose
in the energetic two-body decays of the $B$ meson, as we have elaborated
before (see the end of Sec.~IV). It thus appears to us that the observed large 
decay rates of $B^\pm\to\omega K^\pm$ is attributed to other mechanisms
rather than to a very large value of $\nc$.

   So far we have neglected three effects in the consideration of 
$B^\pm\to\omega K^\pm$: $W$-annihilation, space-like
penguin diagrams and final-state interactions (FSI); all of them are difficult
to estimate. In order to understand why ${\cal B}(B^\pm\to\omega \pi^\pm)
\lsim {\cal B}(B^\pm\to\omega K^\pm)$ experimentally, we need a mechanism
which will only enhance the latter. It appears that FSI may play this role.
Since $B^-\to\omega K^-$ involves only a single
isospin amplitude, inelastic scattering will be the dominant effect of FSI. 
For example, $b\to c\bar c s$ and $b\to u\bar us$ modes can mix with each 
other so that the decay $B^-\to \omega K^-$ arises either from $b\to c\bar cs$
or indirectly through $B^-\to D^0D_s^{*-}$ or $D^{*0}D_s^-$ 
(via $b\to c\bar cs$) with a 
rescattering $D^0D_s^{*-}$ (or $D^{*0}D_s^-$) $\to \omega K^-$. For the decay
$B^-\to\omega\pi^-$,  the inelastic scattering
$B^-\to \{D D^{*}\}\,\to\omega\pi^-$ is Cabibbo suppressed.
Therefore, it is possible that $B^-\to\omega K^-$ receives large FSI from
inelastic scattering but $B^-\to\omega\pi^-$ does not. Since $B^0\to\omega 
K^0$ does not receive contributions from $W$-annihilation,
its measurement can be used to test the relative
strength between FSI and annihilation terms. If the branching ratios of
$B^0\to\omega K^0$ and $B^\pm\to\omega K^\pm$ are close, this will imply
the importance of FSI.

\section{Discussions and conclusions}
For a given effective weak Hamiltonian, there are two important issues in
the study of the hadronic matrix elements for nonleptonic decays of heavy
mesons: one is the renormalization scale and scheme dependence of the
matrix element, and the other is the nonfactorizable effect. For the
former, we have emphasized that it is important to first evaluate the 
vertex and penguin corrections to the matrix
element of 4-quark operators at the scale
$\mu$ so that $\la O(\mu)\ra=g(\mu)\la O\ra_{\rm tree}$ and then
apply factorization or any model calculation to $\la O\ra_{\rm tree}$.
The resulting effective coefficients $c_i^{\rm eff}=c_i(\mu)g(\mu)$
are renormalization-scale and -scheme independent. We pointed out that while
$c_{1,2}^{\rm eff}\approx c_{1,2}^{\rm LO}(\mu)$ at $\mu= m_b(m_b)$ for
current-current operators, the real parts of
$c_{3-6}^{\rm eff}$ are about one and half times
larger than the leading-order penguin Wilson coefficients. This means
that to describe the hadronic charmless $B$ decays dominated by penguin 
diagrams, it is necessary and inevitable to take into account the 
penguin corrections to the 4-quark operators.

    Nonfactorizable effects in hadronic matrix elements of $B\to 
PP,~VP$ decays can be parameterized in terms of the effective number
of colors $\nc$ in the so-called generalized factorization scheme; the
deviation of $1/\nc$ from $1/N_c$ $(N_c=3)$ characterizes the nonfactorizable 
effect. We show that, contrary to the common assumption,
$\nc(V+A)$ induced by the $(V-A)(V+A)$ operators $O_{5,6,7,8}$ are
theoretically and experimentally different from $\nc(V-A)$ generated by
the $(V-A)(V-A)$ operators. The CLEO data of $B^\pm\to\omega\pi^\pm$ 
available last
year clearly indicate that $\nc(V-A)$ is favored to be small, $\nc(V-A)<
2.9\,$. This is consistent with the observation that $\nc(V-A)\approx 2$ 
in $B\to D\pi$ decays. Unfortunately, the significance of $B^\pm\to
\omega\pi^\pm$ is reduced in the recent CLEO analysis and only an
upper limit is quoted. Therefore, a measurement of its branching ratio is
urgently needed. In analogue to the class-III $B\to D\pi$ decays, the
interference effect of spectator amplitudes in charged $B$ decays 
$B^-\to\pi^-\pi^0,~\rho^-\pi^0,~\pi^-\rho^0$ is sensitive to $\nc(V-A)$;
measurements of them [see (3.23)] will be very useful to pin down the value
of $\nc(V-A)$.

   Contrary to the nonfactorizable effects in spectator-dominated rare
$B$ decays, we found that $\nc(V+A)$ extracted from the penguin-dominated
decay $B^\pm\to\phi K^\pm$ is larger than $\nc(V-A)$.
This means that nonfactorizable effects in tree and 
penguin amplitudes behave differently. It turns out this observation is
the key element for understanding the CLEO measurement of $B\to\eta' K$.
In the conventional way of treating $\nc(V+A)$ and $\nc(V-A)$ in the same
manner, the branching ratio of $B^\pm\to\eta' K^\pm$ after including
the anomaly effect in the matrix element $\la\eta' |\bar s\gamma_5s|0\ra$ 
is naively only of order $1\times 10^{-5}$. The running strange quark mass 
at $\mu=m_b$ and SU(3) breaking in the decay constants $f_8$ 
and $f_0$ will enhance ${\cal B}(B\to\eta' K)$ to the order of 
$(2-3)\times 10^{-5}$ with $f_{\eta'}^c
=-6$ MeV. This is still lower than the central value of the CLEO 
measurements. Also, the charm content of the $\eta'$ is not welcome for
explaining the decay rate of $B\to\eta' K$ at small values of $1/\nc$.
We showed that the fact that $\nc(V+A)>\nc(V-A)\approx 2$ will 
substantially enhance the branching ratio of $B^\pm\to\eta' K^\pm$ to 
$(3.7-5)\times 10^{-5}$ at $1/\nc(V+A)\leq 0.2$. 
Unlike the previous analysis, the small charm content of the $\eta'$ is 
now always in the right direction for enhancement irrespective
of the values of $1/\nc(V+A)$. The predicted branching ratio of $B^0\to\eta'
K^0$ is in good agreement with experiment and the calculation of $B^\pm\to
\eta' K^\pm$ is compatible with the data. For a slightly enhanced $f_{\eta'}
^c\approx -15$ MeV, as implied by a recent theoretical estimate, we found
that the agreement of the predicted branching ratio for $B\to\eta' K$ 
with experiment is very impressive. It is thus important to pin down
the decay constant $f_{\eta'}^c$, recalling that the commonly used value
$|f_{\eta'}^c|=6$ MeV is
extracted from experiment within the nonrelativistic quark model
framework. We conclude that no new mechanism in the Standard
Model or new physics beyond the Standard Model is needed to explain
$B\to\eta' K$. 
We have also analyzed charmless $B$ decays into 
the $\eta'$ and $\eta$ in some detail. The branching ratios of the 
spectator-dominated decays $B\to\eta^{(')}\pi,~\eta^{(')}\rho$ were
largely overestimated in the previous analysis because
the matrix element $\la\eta^{(')}|\bar u\gamma_5u|0\ra$ was not 
evaluated correctly before.

    Although the CLEO measurements of hadronic charmless $B$ decays are
satisfactorily explained in the present framework, we found that it is 
difficult to understand the experimental observation that $\Gamma(B^\pm
\to\omega\pi^\pm)\lsim\Gamma(B^\pm\to\omega K^\pm)$. The calculated
branching ratio of $B^\pm\to\omega K^\pm$ is too small compared to
experiment. We conjecture that final-state interactions via inelastic
scattering may contribute in a sizable way to $B^\pm\to\omega K^\pm$,
but are negligible for $B^\pm\to\omega\pi^\pm$ due to the Cabibbo-angle
suppression. Clearly this decay mode deserves further serious investigation 
and a measurement of the neutral decay mode $B^0\to\omega K^0$ will be
very useful to clarify the issue.

   Under the factorization hypothesis, the decays $B\to\phi K$ and 
$B\to\phi K^*$ should have almost the same branching ratios, a prediction not
borne out by current data. Therefore, it is crucial to measure the charged
and neutral decay modes of $B\to\phi(K,K^*)$ in order to see if
the generalized factorization approach is applicbale to $B\to\phi K^*$ decay.

   To conclude, based on the available CLEO data on hadronic charmless
two-body decays of the $B$ meson, we have shown that the 
nonfactorizable effect induced by the $(V-A)(V+A)$ operators is different
from that generated by the $(V-A)(V-A)$ operators. This is the key
element for explaining the CLEO measurement of $B\to\eta' K$.

\vskip 3cm
\acknowledgements

This work is supported in part by the National Science Council of
the Republic of China under Grants NSC87-2112-M001-048 and 
NSC87-2112-M006-018. We are very grateful
to J. G. Smith and X. G. He for helpful discussions and for reading 
the manuscript.

\newcommand{\bi}{\bibitem}

\pagebreak

\centerline{\bf Figure Captions}
\vskip 10mm
\noindent {\bf Fig.~1.}~~ The branching ratio of $B^\pm\to\omega \pi^\pm$ vs
       $1/\nc$. The solid and dashed curves are for $\eta=0.34$, $\rho=0.16$ 
    and $\eta=0.35$, $\rho=-0.12$, respectively. The solid thick lines are
    the CLEO measurements with one sigma errors.
\vskip 3mm

\noindent {\bf Fig.~2.}~~ Same as Fig.~1 except that the branching ratio is
 plotted against $1/\nc(V+A)$ with $\nc(V-A)$ being fixed at the value of 2.
\vskip 3mm

\noindent{\bf Fig.~3.}~~ Same as Fig.~1 except for $B^\pm\to\pi^\pm\pi^0$. The
     thick dotted line is the CLEO upper limit [see (3.20)].
\vskip 3mm

\noindent{\bf Fig.~4.}~~ The branching ratio of $B^\pm\to\phi K^\pm$ vs 
  $1/\nc$ for $\eta=0.34$ and $\rho=0.16$. The dotted curve is for 
   $\nc(V+A)=\nc(V-A)=\nc$ and the solid curve is the branching ratio
   against $1/\nc(V+A)$ with $\nc(V-A)$ being fixed to be 2. The
    solid thick line is the CLEO upper limit.
\vskip 3mm

\noindent{\bf Fig.~5.}~~ Same as Fig.~1 except for $B\to\phi K^*$.
\vskip 3mm

\noindent {\bf Fig.~6.}~~ The branching ratio of $B^\pm\to\eta' K^\pm$ as a
   function of $1/\nc$ for $\eta=0.34$ and $\rho=0.16$.
   The charm content of the $\eta'$ with $f_{\eta'}^c=-6\,{\rm MeV}$ 
   contributes to the solid curves, but not to the dotted curves.
   The lower set of solid and dotted curves takes
   into account the anomaly contribution to $\la\eta'|\bar s\gamma_5s|0\ra$
   [see Eq.~(5.10)], whereas the upper set does not. The solid thick lines
   are the CLEO measurements with one sigma errors.
\vskip 3mm

\noindent {\bf Fig.~7.}~~ Same as Fig.~6 except that the branching ratio is
    plotted against $1/\nc(V+A)$ with $\nc(V-A)$ being fixed at the value of
    2. The anomaly contribution to $\la\eta'|\bar s\gamma_5s
    |0\ra$ is included.
\vskip 3mm

\noindent {\bf Fig.~8.}~~ Same as Fig.~7 except for $B^0\to\eta' K^0$.
\vskip 3mm

\noindent {\bf Fig.~9.}~~ The branching ratio of $B^\pm\to\omega K^\pm$ vs
    $1/\nc(V+A)$ with $\nc(V-A)$ being fixed to be 2. 
    The solid and dashed curves are for $\eta=0.34$, $\rho=0.16$ 
    and $\eta=0.35$, $\rho=-0.12$, respectively. The solid thick lines are
    the CLEO measurements with one sigma errors.
\vskip 3mm

\noindent {\bf Fig.~10.}~~ Same as Fig.~9 except that $\nc(V+A)=\nc(V-A)=\nc$.

\begin{figure}[ht]
\vspace{2cm}
    \centerline{\psfig{figure=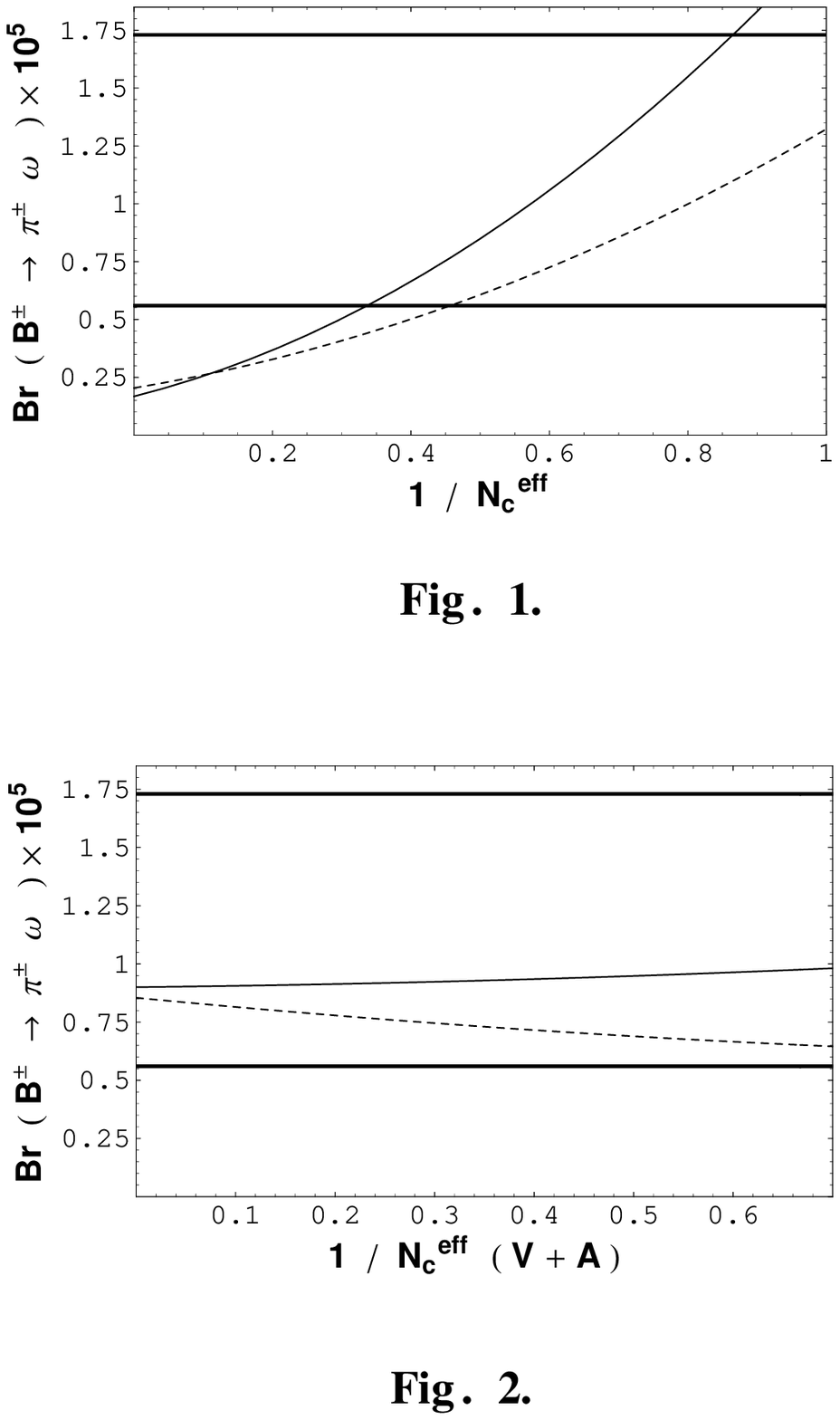,width=11.0cm,height=7.0cm}}
\vspace{4cm}
\end{figure}

\begin{figure}
\vspace{2cm}
    \centerline{\psfig{figure=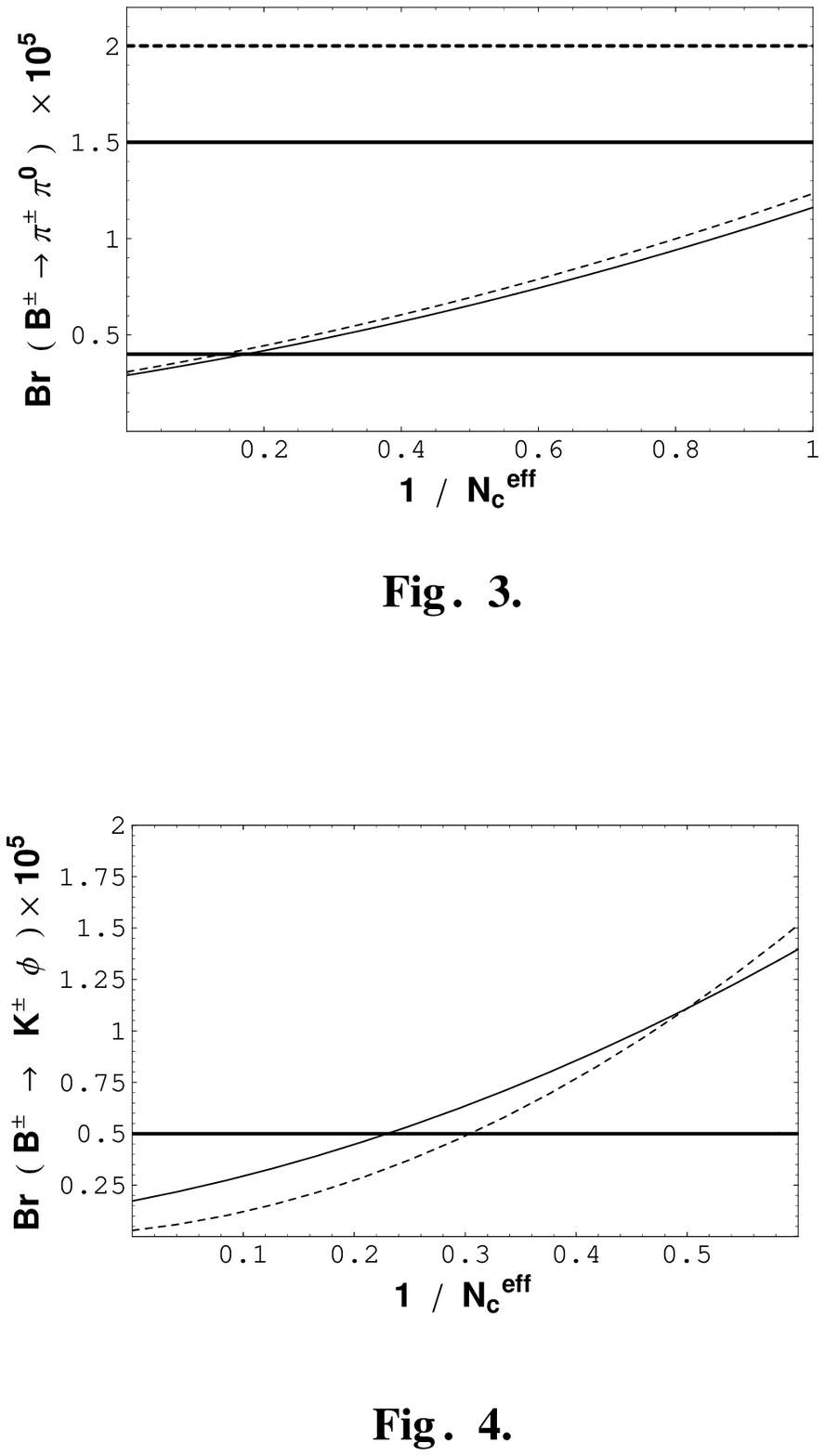,width=11cm,height=7.0cm}}
\vspace{4cm}
\end{figure}

\begin{figure}
\vspace{2cm}
    \centerline{\psfig{figure=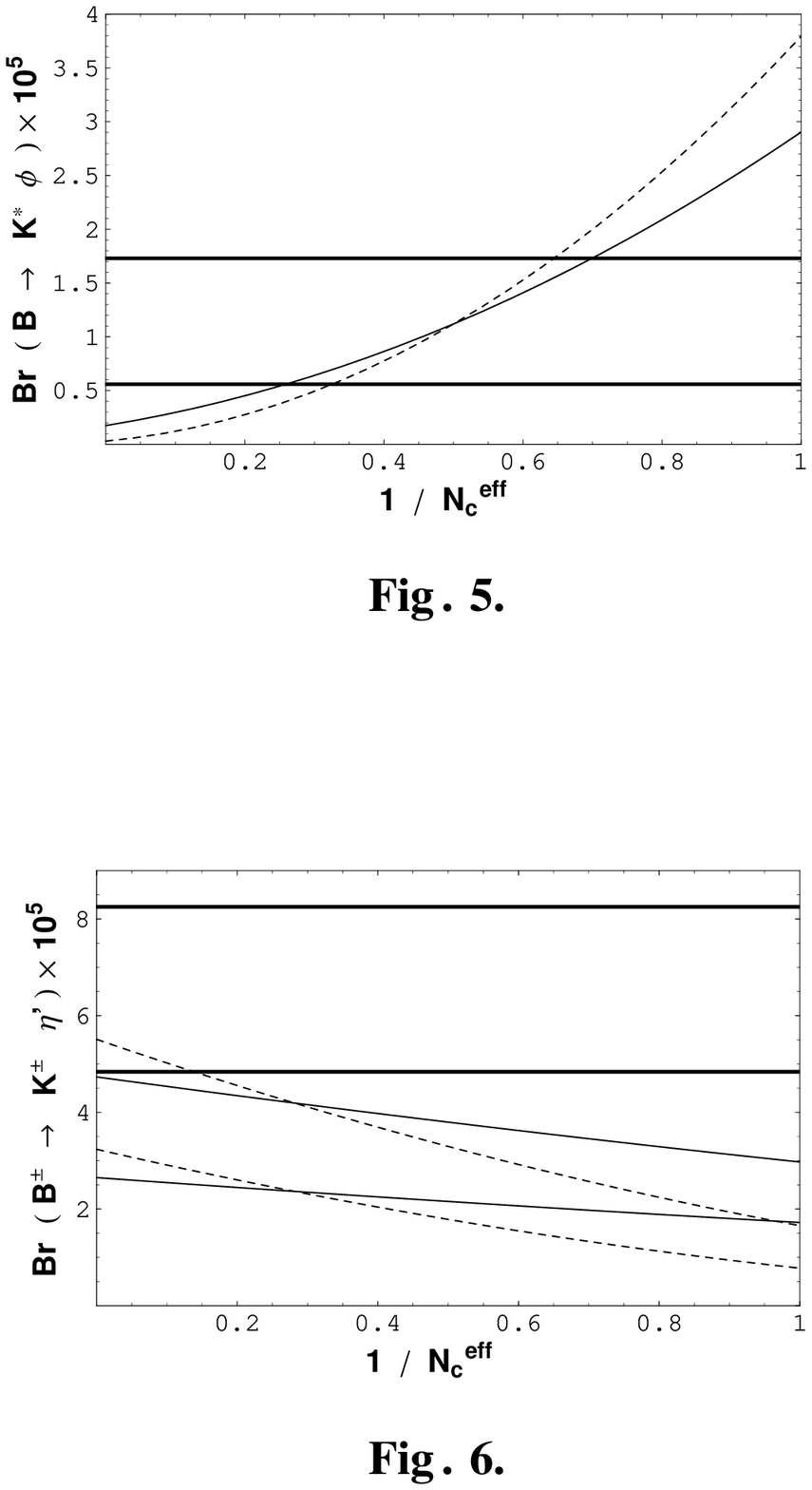,width=11cm,height=7.0cm}}
\vspace{4cm}
\end{figure}

\begin{figure}
\vspace{2cm}
    \centerline{\psfig{figure=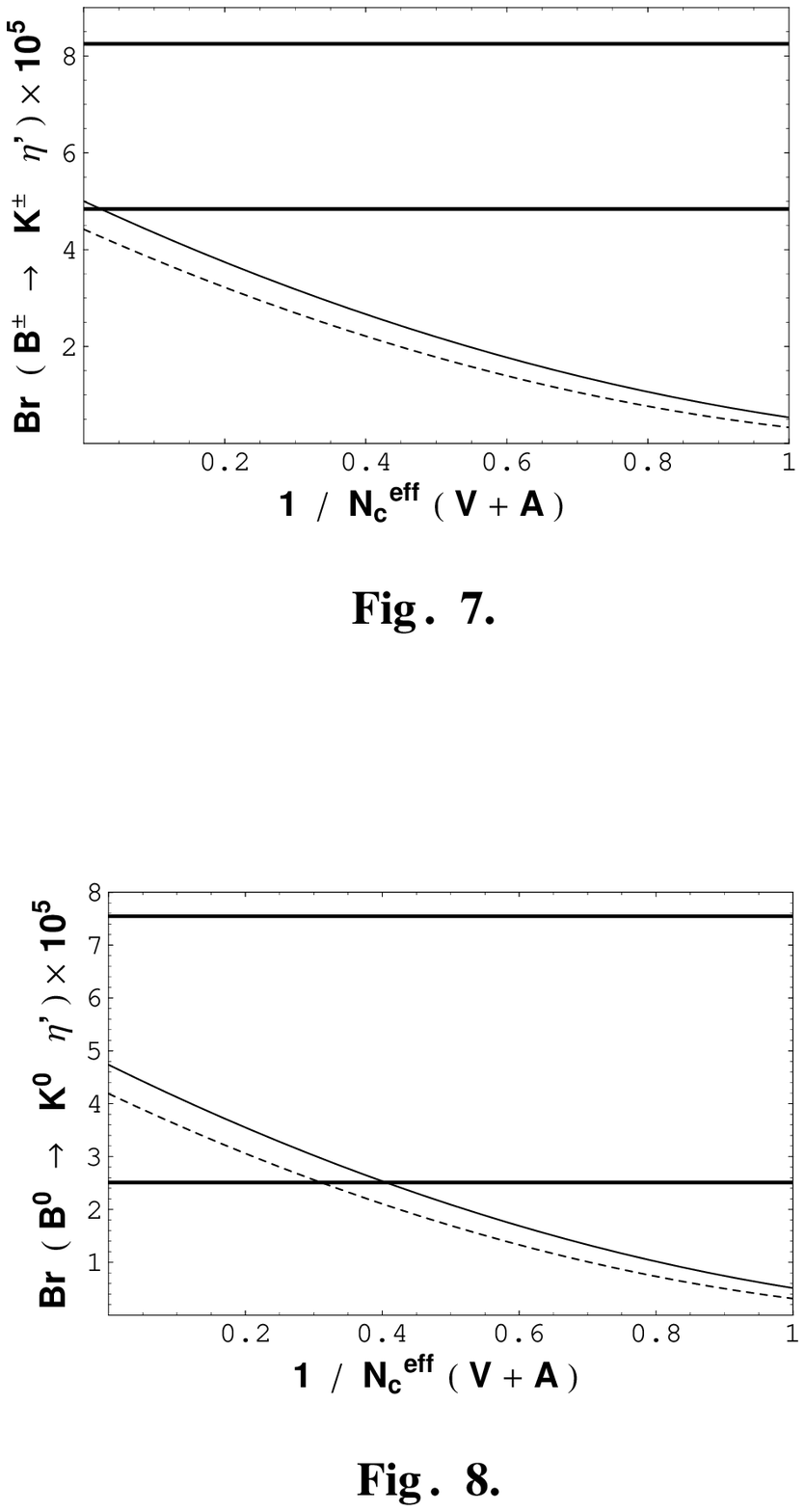,width=11.0cm,height=7.0cm}}
\vspace{4cm}
\end{figure}

\begin{figure}
\vspace{2cm}
    \centerline{\psfig{figure=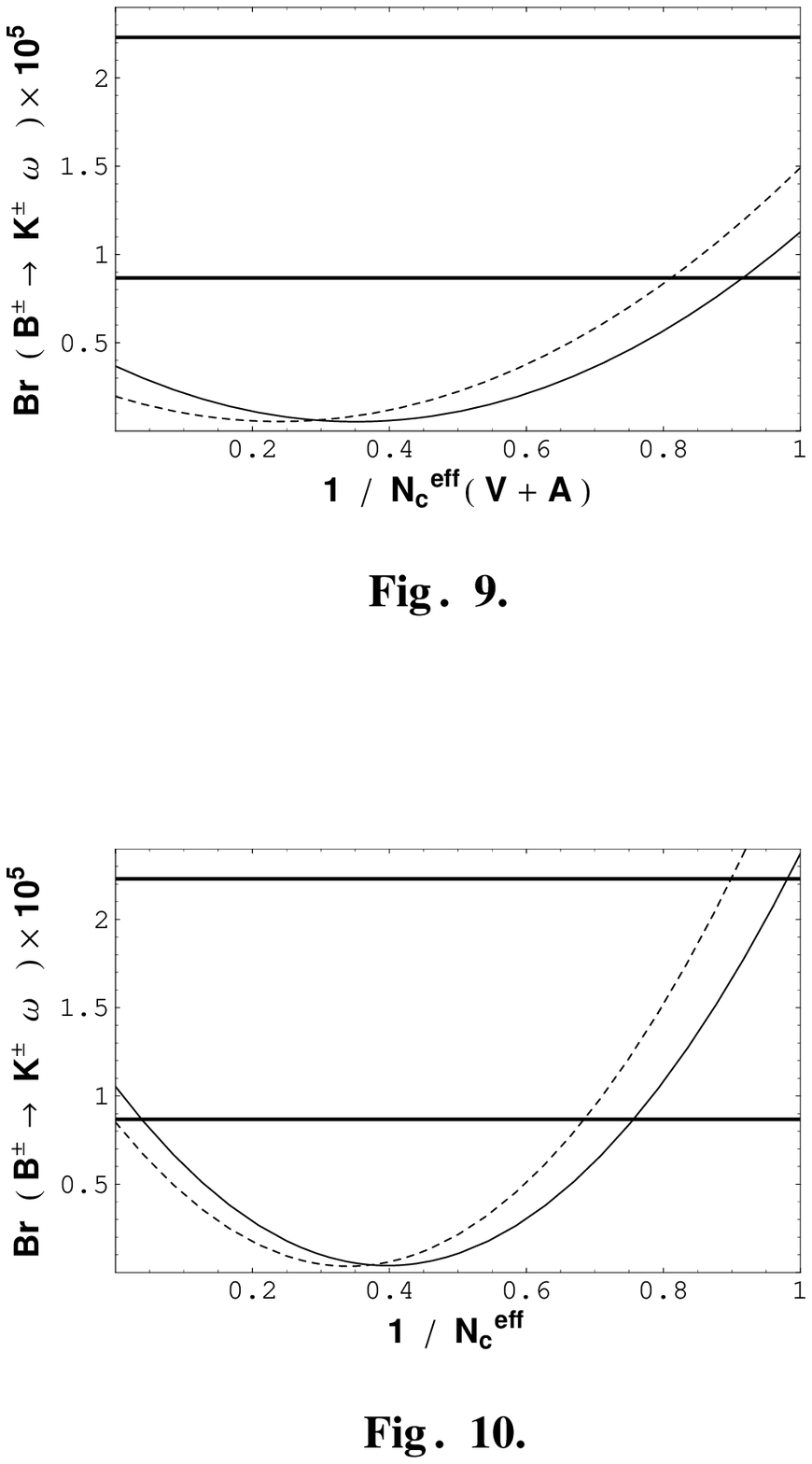,width=11.0cm,height=7.0cm}}
\vspace{4cm}
\end{figure}

\end{document}